\documentclass[12pt]{revtex4}
\usepackage{amsmath}
\usepackage{amsfonts}
\usepackage{amsbsy}
\usepackage{lscape} 
\usepackage{color}
\usepackage{graphicx,epsfig}
\usepackage[english]{babel}
\usepackage{latexsym}
\usepackage{amssymb}
\newcommand{\slashed}[1]{\hbox{{$#1$}\llap{$/$}}}

\begin{document}

\begin{titlepage}
\renewcommand{\thefootnote}{\fnsymbol{footnote}}

\begin{flushright}
FTPI-MINN-05/07\\
UMN-TH-2348/05\\
hep-ph/0505029
\end{flushright}
\vskip -1cm
\begin{center}
\vspace{0.5cm}

\large {\bf Lorentz Violating Supersymmetric Quantum Electrodynamics}\\[3mm] 
  
\vspace*{0.5cm}
\normalsize
{\bf Pavel A. Bolokhov}$^{1,3}$, ~{\bf Stefan Groot Nibbelink}$^{2}$
\ and
{\bf Maxim Pospelov}$^{1,4,5}$

\vspace*{0.5cm}
$^{1}$ {\it Department of Physics and Astronomy,
University of Victoria, Victoria,\\ BC, V8P 1A1, Canada}\\
$^{2}${\it William I.\ Fine Theoretical Physics Institute,
University of Minnesota,\\ Minneapolis, MN 55455, USA}\\
$^{3}${\it Theoretical Physics Department, 
St.\ Petersburg State University, Ulyanovskaya 1,\\
Peterhof, St.\ Petersburg, 198504, Russia}\\
$^{4}$ {\it Perimeter Institute, 31 Caroline Street North,
Waterloo, ON,  N2J 2W9,
Canada}\\
$^{5}$ {\it Department of Physics,
 University of Guelph,
 Guelph, ON,  N1G 2W1, Canada}
 \end{center}

\centerline{\large\bf Abstract}

The theory of Supersymmetric Quantum Electrodynamics is extended by
interactions with external vector and tensor backgrounds, that are   
assumed to be generated by some Lorentz-violating (LV) dynamics at an 
ultraviolet scale perhaps related to the Planck scale. Exact supersymmetry
requires that such interactions correspond to LV operators of
dimension five or higher, providing a solution to the naturalness
problem in the LV sector. We classify all dimension five and six LV
operators, analyze their properties at the quantum level and describe
observational consequences of LV in this theory. 
We show that LV operators do not induce destabilizing D-terms,
gauge anomaly and the Chern-Simons term for photons. We calculate 
the renormalization group evolution of dimension five LV operators and
their mixing with dimension three LV operators, controlled by the
scale of the soft-breaking masses. Dimension five LV operators are
constrained by low-energy precision measurements at 
$10^{-10}-10^{-5}$ level in units of the inverse Planck scale, while
the Planck-scale suppressed dimension six LV operators are allowed by
observational data.

\end{titlepage}

\newpage

\setcounter{footnote}{0}
\setcounter{equation}{0}

\section{Introduction}
\label{Intro}

There are many known examples in the history of physics when a 
symmetry of nature, which was assumed to be exact, has fallen under
experimental scrutiny. The study of the consequences of such breaking
has often provided important insights into the dynamics at
high-energy scales. This was exemplified by the weak-scale dynamics of
the Standard Model (SM) through the search and discovery of P and CP 
violations. Lorentz symmetry is used as a crucial ingredient in the
construction of fundamental theories of nature. Even though no
breakdown of this symmetry has been observed to date, there has been a
growing interest in Lorentz Violation (LV) because the precision 
tests of Lorentz symmetry can provide an important window into the
physics far beyond the electroweak scale
\cite{Kost1,CG,Jacobsonreview,PhysToday,Sigl:2004cq,Mattingly:2005re}.

The recently intensified interest in LV theories is stimulated by
several seemingly unrelated motives. Firstly, a combination of different sets of 
cosmological data indicates that the dominant component of the energy
density of the universe is dark energy. It can either be ascribed
to a cosmological constant or to an energy density associated with a
new infrared degree of freedom, such as {\em e.g.}\ an ultra-light
scalar field (quintessence). The time evolution of quintessence
creates a preferred frame, which could in principle be detected as a
LV background, provided that it couples to the SM. Secondly, low-energy
limits of string theory contain a number of (nearly) massless fields,
some of which carry open Lorentz indices. The well-studied 
example of an antisymmetric field background $B_{\mu\nu}$ on a brane
(for a review see \cite{DN}) leads to an effective violation of Lorentz
invariance. Thirdly, there have been a number of conjectures that a
theory of quantum gravity could manifest itself at lower energies
through LV modifications of particle dispersion relations 
(see, {\em e.g.}\ \cite{lcq,Vucetich:2005ra} and references
therein). Although such 
conjectures are undoubtedly speculative, if true, they would provide a
powerful tool to probe ultra-short distances via LV physics.  Direct
experimental constraints on modifications of dispersion relations come
from astrophysical processes \cite{CFJ,AmC,Ted1,GK,Kost2,Bertolami:1999da,Sarkar,Bertolami:2003qs} and
terrestrial clock comparison experiments \cite{clock1,clock2,Vuc,MP:}. 
In both cases the typical sensitivity to these operators is at the
$10^{-5}/M_{\rm Pl}$ level. This creates a definite problem for those
theories that predict Planck mass suppressed $\sim 1/M_{\rm Pl}$
effects.

In an effective field theory framework, the breakdown of Lorentz 
symmetry can be described by the presence of external tensors, which
are generated by some unspecified dynamics, coupled to SM operators. 
It is useful to characterize such operators by powers of increasing
dimension, as it gives an indication to the possible scaling of LV
effects with the ultraviolet (UV) scale $M$, which might be related to $M_{\rm Pl}$. In
quantum electrodynamics (QED), the generic expansion in terms of the
gauge invariant operators starts at dimension three (see {\em e.g.}\ 
\cite{Kost1}): 
\begin{equation}
{\cal L}_{\rm QED}^{(3)} =
~-~a_\mu\,  \bar \Psi \gamma_\mu \Psi
~-~ b_\mu\,  \bar \Psi \gamma^\mu \gamma_5 \Psi 
~-~ \frac{1}{2}H_{\mu\nu}\bar \Psi \sigma^{\mu\nu} \Psi
~-~ k_\mu\,  
\epsilon^{\mu\nu\kappa\lambda} A_\nu \partial_\kappa A_\lambda~.
\label{LVqed}
\end{equation}
Here $\psi$ is a Dirac spinor describing the electron and $A_\mu$ is the
electromagnetic vector-potential. The external vector and
anti-symmetric tensor backgrounds, $a_{\mu}$, $b_\mu$, $k_\mu$ and
$H_{\mu\nu}$ define a preferred frame, and therefore break Lorentz
invariance. The coupling to the vector current, 
$a_\mu \bar \Psi \gamma^\mu \Psi$, can be removed by introducing a
space-time dependent phase for the electron. The last term in the
Lagrangian \eqref{LVqed}, the Chern-Simons term, is gauge invariant up
to a total derivative, which can be neglected.

At this dimension three level there is a problem in ascribing LV to UV
dynamics. From simple dimensional counting  one would expect the
external vectors and tensors, $a_\mu,~b_\mu,\ldots$ to be on the order
of the UV scale $M$, and therefore LV would be very large, which is clearly
inadmissible. For example, a Higgs mechanism resulting in a
condensation of a vector field $V_{\mu}\sim M n_\mu$ (where $n_\mu$ is a 
"unit" vector \cite{Kostelecky:1989jw}) creates disastrous
consequences when coupled to a non-conserved current, 
{\em i.e.}\ the axial current $\bar \Psi \gamma_\mu\gamma_5 \Psi$.  
One may hope that operators of dimension three and four are forbidden by 
some symmetry arguments or tuned to be small, so that LV effects 
first appear at dimension five (or higher) level \cite{MP:}. 
However, such hopes are typically shattered by quantum corrections, which
lead to dimensional transmutation of a higher dimensional operator into
a lower dimensional one with a quadratically divergent coefficient: 
\begin{equation}
[LV]_{\rm dim~3} ~~\sim~~ ({\rm loop~factor}) \, 
\;\Lambda_{UV}^2\;
\times\; [LV]_{{\rm dim}~5}~. 
\label{transmute}
\end{equation}
Here $[LV]_{\rm dim~3(5)}$ represent generic LV operators of dimension
three and five, respectively. If the UV cutoff scale $\Lambda_{UV}$ is
of the order of $M$, huge dimension three operators are generated. In
that case all higher dimensional operators would have to be tuned,
leaving no room for LV interactions. This naturalness problem of LV
physics can be avoided if these quadratic divergences are suppressed
by certain symmetry arguments. In Ref.~\cite{MP:} it was shown that
dimension five LV operators coupled to fully symmetric three-index
traceless tensors are protected against developing quadratic loop 
divergences. But this solves the naturalness problem only partially, as
this does not provide an argument as to why dimension three and four
operators cannot be induced at tree level, and why they have to be
tuned by hand to experimentally acceptable values.

A recent paper \cite{GrootNibbelink:2004za} proposed that
supersymmetry (SUSY) could provide a powerful selection rule on
admissible forms of LV interactions. In particular, it has been shown
that in the Minimal Supersymmetric Standard Model (MSSM) the
requirements of SUSY and gauge invariance restrict LV operators to be
of dimension five or higher. Therefore SUSY solves the naturalness
problem of LV physics. Once SUSY is softly broken, the quadratic UV
divergences are effectively stabilized at  the supersymmetric
threshold. Hence this might lead to a solution of the question of 
why the lower dimensional LV operators are so much suppressed as
compared to their natural scale.

An explicit example of how SUSY restricts possible LV interactions and
leads to dramatic numerical changes in the predicted observables is
provided by non-commutative field theories. The non-commutative
background tensor $\theta_{\mu\nu}$, entering the Moyal product, has
the canonical dimension $-2$, and therefore the scale of non-commutativity,
$\Lambda_{NC} \sim(\theta)^{-1/2}$, gives a natural UV scale. 
As a result, linearizing the action in $\theta$ is justified, as long as the
characteristic momenta are much smaller than $\Lambda_{NC}$. 
This expansion leads to a set of dimension six operators, which at the
tree level induce interactions between particle spins and the 
$\theta_{\mu\nu}$ background \cite{MPR1} with effective 
$H_{\mu\nu}$ in (\ref{LVqed}) given by 
$H_{\mu\nu} \sim\Lambda_{IR}^3\theta_{\mu\nu}$. 
Here $ \Lambda_{IR}$ is the relevant infrared scale, such as 
$\Lambda_{QCD}$ in the case of hadrons. However, it has been shown
that loop effects in non-commutative field theories lead to
quadratically divergent integrals \cite{UCSC}: 
$H_{\mu\nu} \sim \Lambda_{IR}\Lambda_{UV}^2\theta_{\mu\nu}$. 
This essentially invalidates the expansion in terms of $\theta_{\mu\nu}$. 
If the cutoff scale is very high, {\em e.g.}\ comparable to $\Lambda_{NC}$, 
the resulting spin anisotropy is large and certainly excluded by
experiment. However, this conclusion is premature as  
one can argue that the operator $\bar q \sigma_{\mu\nu} q$ 
is incompatible with SUSY \cite{MPR2} and thus should not be induced 
in the domain of the loop momenta higher than the SUSY breaking. 
This means that the cutoff essentially coincides with the 
energy splitting between fermions and bosons, {\em i.e.}\ 
$\Lambda_{UV} \sim m_{\rm soft}$. This has been confirmed by an
explicit two-loop calculation in the framework of  non-commutative
supersymmetric QED \cite{WMC2}. With the quadratic divergences 
stabilized at $m_{\rm soft}\sim 1~{\rm TeV}$,
the Planck scale non-commutativity is safely within the experimental 
bounds. This example illustrates that the existence of SUSY can be
important for understanding the actual size of the expected LV effects. 
Another example as to how SUSY can protect against quadratic
divergences in a LV theory has been given recently in \cite{Jain:2005as}.

The purpose of this work is to analyze in detail LV operators in 
supersymmetric quantum electrodynamics (SQED), to prove 
the absence of the naturalness problem in the LV sector, and to derive
phenomenological  constraints on the LV parameters in SQED.  
We see this as a first step towards the phenomenological analysis of the full
LV MSSM.  Following Ref.~\cite{GrootNibbelink:2004za}, we
parametrize all dimension five operators of LV SQED by three vector 
$N^{\mu}$, $N^{\mu}_+$ and $N^{\mu}_-$ and one irreducible 
rank three tensor $T^{\mu\,\nu\lambda}$ backgrounds. 
These vector and tensor backgrounds enter in the LV
operators composed  of a vector superfield (containing photons and
photinos) and chiral superfields (corresponding to left- and
right-handed (s)electrons), respectively. 
We introduce these CPT-violating operators in the superfield
formalism, and then derive their component form. We also classify
dimension six CPT-conserving LV operators in superspace notations. 
We observe that by using the equations of
motion (EOM's) some parts of dimension five operators can be 
reduced to dimension three LV operators. The relation  
between them
$[LV]_{\rm dim~3} \sim m_e^2\, [LV]_{\rm dim~5}$ 
is controlled by the electron mass $m_e$.

The main emphasis of our study is on the quantum effects. We show that
even in the presence of SUSY LV operators no destabilizing quadratically
divergent D-terms ever arise. We prove that gauge anomalies are
not affected by the presence of these LV operators. This analysis
essentially implies that the Chern-Simons (CS) term cannot arise from quantum
corrections. We derive the renormalization group (RG) evolution for the
LV operators, showing explicitly that only the logarithmic divergences
arise in the limit of exact SUSY. We solve the one-loop
renormalization group equations (RGE's)  to obtain the low-energy
values of LV parameters in terms of their values at the UV scale
$M$. Then we investigate the consequences of SUSY breaking for
LV operators by introducing the soft-breaking masses for superpartners
of electrons. Dimension three LV operators can now be induced by
dimensional transmutation: 
$[LV]_{\rm dim~3} \sim m_{\rm soft}^2 \, [LV]_{\rm dim~5}$. 
Although a loop effect, this constitutes a dramatic enhancement
compared to 
the case with unbroken SUSY, as $ m_{\rm soft}^2/m_e^2 > 10^{10}$.  
One might expect that similar quantum corrections could induce a CS
term (the last operator in \eqref{LVqed}) once SUSY is
broken. However, our analysis rules out this possibility.

We investigate phenomenological consequences of LV in the framework of
softly-broken SQED. The strongest constraints on the LV parameters are
due to the (non)observation of anomalous spin precession
around directions defined by the LV background vectors $N^\mu,
N_+^\mu$ and $N_-^\mu$. Another constraint comes from the
comparison of the anomalous magnetic moments of electrons and  
positrons. It is important to note that all constraints obtained in this 
work are laboratory constraints, as astrophysical and cosmological 
searches of LV are not sensitive to LV effects in SQED.

We present our results in the following order. Section \ref{LVoperators} 
introduces the LV operators and backgrounds at dimension five and six
levels. Section \ref{quantum} investigates various quantum corrections
to LV operators under the assumption of exact SUSY. Subsection
\ref{noDterm} shows that no dangerous quadratically divergent D-terms
arise. Subsection \ref{noAnomaly} explains that no novel gauge
anomalies can ever appear due to LV operators, and consequently that
a SUSY CS term is ruled out. Finally, subsection \ref{RGEvolution}
addresses the running of LV operators of dimension five. Section
\ref{InducedDim3} studies the consequences of soft SUSY breakdown for
the LV sector, and derives RGE's for the induced dimension three LV
operators. In subsection \ref{SB_gauge_sector} we argue that even when
SUSY is broken no CS term is generated. In section \ref{Phenomenology}
we study the phenomenology of the model, and obtain various
predictions for relevant LV observables.  We reach our conclusions in
section \ref{conclusion}.

\section{LV operators in SQED} 
\label{LVoperators}

Supersymmetric Quantum Electrodynamics (SQED) is described
by two chiral superfields $ \Phi_+ $ and $ \Phi_- $, that 
are oppositely charged under a U(1) gauge superfield $ V $: 
\begin{eqnarray}
\mathcal{L}_{\mathrm SQED} & ~=~
&
\int d^4\theta\, \Big(
   \overline{\Phi}_+ e^{2eV} \Phi_+ ~+~
   \overline{\Phi}_- e^{-2eV} {\Phi}_-  \Big) \\
\label{SQED}
&& +~   
\int d^2\theta\, \Big( \frac{1}{4}\,  WW ~+~m_e\, \Phi_-\Phi_+ \Big) ~+~
\int d^2\overline{\theta}\, 
\Big( \frac{1}{4}\, \overline{W}\,\overline{W} ~+~ 
\overline{m}_e\, \overline{\Phi}_+\overline{\Phi}_- \Big)~.
\nonumber
\end{eqnarray}
Here $ W_\alpha = - \frac{1}{4} \overline{D}{}^2 D_\alpha\, {V} $ 
is the super gauge invariant expression for the field strength. 
Throughout this paper, we use predominantly Wess and
Bagger notations \cite{Wess:1992cp}. The fermionic components of
superfields $ \Phi_+ $ and $ \Phi_- $ correspond to the left-handed
electron and right-handed charge-conjugated electron fields. With a   
slight abuse of the language, we call them the electron and positron 
superfields, or just the electron and the positron for brevity. We
define the charge of electron as $ e = - | e | $. 
Finally, $m_e$ denotes the (complex) electron mass.

LV extensions of SQED can be constructed as a set of effective
operators containing the superfields $\Phi_-$, $\Phi_+$,
gauge covariant derivatives $ \nabla_\alpha $, 
$ \overline{\nabla}_{\dot\alpha} $ and {\em arbitrary} constant tensor
coefficients with Lorentz indices  that specify the
breakdown of Lorentz invariance \cite{GrootNibbelink:2004za}. 
The general rules according to which LV operators should be
constructed are listed in Ref.~\cite{MP:}. 
Within the context of SQED, however, we impose additional
requirements related to supersymmetry.
In this work we require that all LV operators be  
\begin{itemize}
\item supersymmetric, 
\item local super gauge invariant with chiral gauge parameters, 
\item have local component expressions. 
\end{itemize}
Let us explain these conditions in more detail. 

First of all,
by having supersymmetry we mean that the sub-algebra 
\begin{equation}
\{ Q_\alpha, \overline{Q}_{\dot\alpha} \} ~=~ 
2\, \sigma^\mu_{\alpha\dot{\alpha}} \, P_\mu
\end{equation}
of the $N=1$ super Poincar\'e algebra remains unbroken. (LV Theories with
higher amounts of SUSY coming from extra dimensions have also
been investigated  \cite{Ney:2004tn,Ney:2005gb,Ney:2005wc}).
If we assume that the breaking of the Lorentz symmetry is 
{\em spontaneous}, we are guaranteed that
$\sigma^\mu_{\alpha\dot{\alpha}}$ represent the standard Pauli
matrices. However, if the breaking of Lorentz symmetry is {\em
explicit} from the outset of the theory, these objects are simply 
structure coefficients parameterizing this supersymmetry algebra. (In
this work we do not pursue this possibility further. Possible
modifications of superalgebra by LV parameters has been discussed in
Ref.~\cite{Berger:,Berger:2003ay}.) This assumption allows us to perform our
analysis using conventional superspace.

The requirement of having a local component expression allows for a
conventional effective field theory interpretation of the Lagrangians that
we obtain. However, the locality of the component Lagrangian
does not necessarily imply that the superspace
expression of a given Lagrangian appears local 
\footnote{We would like to thank N.\ Arkani-Hamed for this comment.}. 
For example, the electron mass term can be written in a seemingly
non-local way:
\(
\int d^4 \theta\, m_e \, \Phi_- D^2/(- 4\Box) \Phi_+ + \text{h.c.}.
\)
Finally, we require that LV operators preserve the standard local
super gauge transformations  
\begin{equation}
\Phi_\pm ~\rightarrow~ e^{\mp 2 e \Lambda} \, \Phi_\pm~, 
\qquad 
\overline{\Phi}_\pm ~\rightarrow~ e^{\mp 2 e \overline{\Lambda}} \, 
\overline{\Phi}_\pm~, 
\qquad 
V ~\rightarrow~ \Lambda ~+~ \overline{\Lambda}~, 
\label{Gauge}
\end{equation} 
with a chiral parameter $\Lambda$. In particular, we do not allow for
non-local or non-chiral extensions of the gauge transformations that
seem to be required by non-commutative SUSY
\cite{Putz:2002ib,Mikulovic:2003sq}.

As was shown in \cite{GrootNibbelink:2004za}, these conditions 
combined impose strong restrictions on the number of LV terms of
a specific mass dimension one can construct: no dimension three or four
LV operators can be written down within the context of the MSSM. 
Here we do not repeat all the arguments leading to this general claim,
but simply illustrate the underlying philosophy by showing that the CS
term (the last interaction in \eqref{LVqed}) does not have a SUSY
extension satisfying all three conditions stated above.

The CS term is a dimension three operator that is bilinear in the
gauge field and proportional to an external vector. Therefore the
local superspace extension of it can be represented as 
\begin{equation}
{\cal L}_{\rm SCS}^{\rm local} ~=~ 
\frac 12 \, k^\mu \int d^4 \theta \, 
 \overline{\sigma}^{\dot{\alpha} \alpha}_\mu\, 
V [ D_\alpha, \overline{D}_{\dot\alpha}] V 
~=~
k_\mu \, \Big( 
- \,\epsilon^{\mu\lambda\nu\rho} \, 
A_\lambda \partial_\nu A_\rho
~+~ 
2\,  A^\mu D
~+~
2\, \overline{\lambda}\,\overline{\sigma}^\mu \lambda
\Big)~. 
\label{locCS}
\end{equation} 
This is the only possible structure, as the insertion of an
anti-commutator  $\{D_\alpha, \overline{D}_{\dot\alpha}\}$ immediately
gives rise to a total spacetime derivative. The component expression 
shows that this operator indeed contains the CS term, which 
is gauge invariant up to a total derivative. However, the
SUSY extension as a whole is not super gauge invariant: 
\begin{equation}
\delta {\cal L}_{\rm SCS}^{\rm local} 
~=~ 2 i\, k^\mu 
\int d^4 \theta \, 
V\, \partial_\mu ( \overline{\, \Lambda} - \Lambda \,)~. 
\label{locCStrans} 
\end{equation} 
Notice that this statement is independent of the Wess-Zumino
gauge, and that even under the restriction of  gauge invariance
under ordinary U(1) transformations ($\Lambda = i \,\alpha$) the
supersymmetric extension (and the $A^\mu D$ term in particular)  of
the CS term fails to be gauge invariant.

These arguments do not show that it is impossible to construct a super
gauge invariant extension of the CS term. Indeed, by inserting the
transversal projector 
$P_V = D^\alpha \overline{D}{}^2 D_\alpha/(-8 \Box)$
we obtain a manifestly super gauge invariant expression 
\begin{equation}
\label{nonlocCS}
{\cal L}_{\rm SCS}^{\rm non-local} ~=~ 
\frac 12 \, k^\mu \int d^4 \theta \, 
 \overline{\sigma}^{\dot{\alpha} \alpha}_\mu\, 
V\, P_V\,  [ D_\alpha, \overline{D}_{\dot\alpha}] V
~=~
k^\mu \int d^4 \theta \, 
\overline{W} \, \overline{\sigma}_\mu \, \frac 1{\Box} W
~.
\end{equation} 
This expression clearly appears to be non-local in superspace, but
the CS term itself is still local. In fact, because the true CS term
in \eqref{locCS} already was gauge invariant, the insertion of $P_V$
did not affect it  at all. However, other terms in the component
expression of \eqref{nonlocCS} are non-local because they contain
$1/\Box$ explicitly.  Hence, as asserted, the CS term does not allow
for a SUSY extension that is super gauge invariant and that has a
local component expression.  Additional discussion of LV due to a CS
term in supersymmetric theories  
can be found in Refs.~\cite{Belich:,Belich:2005js}.

\subsection{CPT-violating dimension five LV operators}
\label{Dim5LV}

There are only three different types of LV operators satisfying the 
above requirements in SQED at the
dimension five level. In this subsection we give their superfield
expressions, while their component forms can be found in Section
\ref{Phenomenology}. The first type is the electron and positron 
superfield operators  
\begin{equation}
\label{LV_matter}
  \mathcal{L}_{\mathrm{LV}}^{\mathrm{matter}} ~=~ 
\frac{1}{M}\,   \int d^4\theta \Big\{ 
N_+^\mu\, \overline{\Phi}_+ e^{2eV} i \nabla_\mu \Phi_+ 
~+~ N_{-}^\mu\, \overline{\Phi}_- e^{-2eV} i \nabla_\mu  {\Phi}_-
                 \Big\}~, 
\end{equation}
which are parameterized by two external real vectors $N_\pm^\mu$.  
The super gauge covariant spacetime derivative 
$\nabla_\mu   =  - \frac{i}{4} 
\bar{\sigma}_\mu^{\dot{\alpha}\alpha}
\{ \nabla_\alpha, \overline{\nabla}_{\dot{\alpha}} \} 
$ 
is defined in terms of the super gauge covariant derivatives 
$\nabla_\alpha$ and $\overline{\nabla}_{\dot{\alpha}}$. Their precise
form depends on the super gauge transformation properties of the
object that they act on. For example, we define 
\begin{equation}
\nabla_\alpha \Psi_\pm ~=~ 
e^{\mp 2eV} \, D_\alpha \, \big( e^{\pm 2eV} \Psi_\pm \big)~, 
\qquad 
\overline{\nabla}_{\dot{\alpha}} \Psi_\pm ~=~
\overline{D}_{\dot{\alpha}} \Psi_\pm~,
\end{equation} 
for generic superfields $\Psi_\pm$, that have the same gauge
transformations as the (chiral) superfields $\Phi_\pm$, see 
\eqref{Gauge}.

For the photon super multiplet we can construct two independent
operators. The first operator is parameterized by a real vector
$N^\mu$. We can give a K\"ahler-like representation of this vector
operator as  
\begin{equation}
\label{LV_gauge}
\mathcal{L}_{\mathrm{LV\ dim\ 5}}^{\mathrm{gauge\ (V)}} ~=~ 
\frac 1M \int d^4\theta \, 
N^\kappa\, \overline{W} \bar{\sigma}_\kappa W~.   
\end{equation}
Using a superspace identity, this operator can also be written as a
superpotential-like term  
\begin{equation}
\label{LV_gauge_Alt}
\mathcal{L}_{\mathrm{LV\ dim\ 5}}^{\mathrm{gauge\ (V)}} ~=~ 
- \, \frac {N_\kappa}{2 M}\,  \epsilon^{\kappa\lambda\mu\nu} \,  
\Big( 
\int d^2\theta\, W \sigma_{\mu\nu} \partial_\lambda W ~+~
\int d^2\bar{\theta}\, \overline{W} \, \bar{\sigma}_{\mu\nu}\, 
\partial_\lambda \overline{W} 
\Big)~.
\end{equation} 
The most general LV superpotential-like term takes the form 
\begin{equation}
\label{LV_gauge_Tterm}
\mathcal{L}_{\mathrm{LV\ dim\ 5}}^{\mathrm{gauge\ (T)}} ~=~ 
\frac 1{4M} 
\int d^2\theta \, T^{\lambda\, \mu\nu} \,
        W \sigma_{\mu\nu} \, \partial_\lambda W  
~+~ \frac 1{4M} 
\int d^2\theta \, \overline{T}^{\lambda\, \mu\nu} \,
        \overline{W} \,\bar{\sigma}_{\mu\nu}\, \partial_\lambda\overline{W}  
~.
\end{equation}
In principle this operator is parameterized by a complex rank-three
tensor $T^{\lambda\,\mu\nu}$, antisymmetric in the last two indices
$(\mu,\nu)$ due to its contraction with $\sigma_{\mu\nu}$. Notice that
$\sigma_{\mu\nu}$ acts as a projector on the imaginary self-dual part
of the tensor since 
\(
\frac{1}{2}\,i\,\epsilon_{\mu\nu}{}^{\rho\sigma}
\sigma_{\rho\sigma} = \sigma_{\mu\nu}. 
\)
This implies that we may take $T^{\lambda\,\mu\nu}$ real. We can
constrain it further by requiring that 
\begin{equation}
T_\mu^{\phantom{\mu}\mu\rho} ~=~ 0~,
 \qquad 
\epsilon_{\kappa\lambda\rho\sigma}\, T^{\lambda\,\rho\sigma}  ~=~  0~.
\end{equation} 
The first condition arises because any trace part of
$T^{\lambda\,\mu\nu}$ of the operator \eqref{LV_gauge_Tterm} 
vanishes, as 
\(
\int d^2\theta \, W\sigma^{\mu\nu} \partial_\nu W + 
\int d^2\theta \, \overline{W} \bar\sigma^{\mu\nu} \partial_\nu
\overline{W} = 0. 
\)
The second condition ensures that the LV due to a vector 
background is entirely accounted for by \eqref{LV_gauge_Alt}.

In a non-Abelian theory, operator \eqref{LV_gauge_Tterm} cannot exist
because the $W_\alpha$'s are not gauge invariant but only gauge covariant.
Thus, to maintain gauge invariance, any derivative acting on 
$ W_\alpha $ has to
be replaced by a corresponding super gauge derivative. 
In particular, a non-Abelian generalization of 
 (\ref{LV_gauge_Tterm}) would have to contain the covariant
derivative $\nabla_\mu$.  
But then the integrand would not be chiral, as 
\begin{equation}
\overline{D}_{\dot\gamma} \nabla_\mu W_\alpha ~=~ 
-i e\, (\epsilon \bar\sigma_\mu)_{\dot\gamma}{}^\beta\, 
\Big( W_\alpha W_\beta ~+~ W_\beta W_\alpha \Big) ~\neq~ 0~. 
\end{equation} 
Therefore, in the non-Abelian case one cannot write down
superpotential-like LV terms for gauge multiplets, and only 
the K\"ahler-like terms \eqref{LV_gauge} are allowed. 

We have listed all possible dimension five operators in SQED
framework. These results have been reported before in the MSSM setting
\cite{GrootNibbelink:2004za}. 
All operators of dimension 5, listed in this section, break CPT invariance. 
The CPT-conserving LV operators start at dimension 6 level. 
For the matter of completeness, we now classify all dimension 
six LV operators compatible with SQED. 
However, our main analysis of quantum loop effects and observational 
implications of LV will be concentrated on the dimension five
operators \eqref{LV_matter}, \eqref{LV_gauge} and \eqref{LV_gauge_Tterm}.

\subsection{CPT-conserving dimension six LV operators}
\label{Dim6}

Let us start by considering possible superpotential-like terms. 
To obtain dimension six operators in the Lagrangian density, 
one has to consider the superpotential at dimension five level, 
{\em i.e.}\ two dimensions higher than the standard mass term $m_e\,
\Phi_-\Phi_+$. Because of the chirality condition  of the
superpotential, gauge invariance and the absence of 
fermionic LV backgrounds,  all possible terms have to be built out of the
(dimension two) operator $\Phi_+ \Phi_-$ and the (dimension three)
operator $W_\alpha W_\beta$, with possible derivative insertions in  
the latter. Omitting all Lorentz-preserving terms in the
superpotential, we arrive at the following LV operator at
dimension six level,  
\begin{gather} 
{\cal L}_{\rm{LV\ dim\ 6}}^{\rm{super}} ~=~ \frac{1}{M^2}
\int d^2 \theta \, 
S^{\mu\nu}\, W \partial_\mu \partial_\nu W 
~+~ \text{h.c.}~. 
\label{LV_dim6_Fterm}
\end{gather}
The dimensionless matrix $S$ is symmetric: 
$S^{\mu\nu} =S^{\nu\mu}$. 
All other possible operators would involve $W \sigma_{\mu\nu} W$ 
which vanishes for a single U(1). 
%
%
As mentioned before, the superpotential term \eqref{LV_dim6_Fterm} 
can be represented as an 
integral over the full superspace by factoring out $-\frac 14
\overline{D}{}^2$. This can be done in various ways leading to
seemingly different expressions for these operators. Since the
superpotential expression above defines these operators uniquely,
there is no need to give full superspace representations of 
these operators here.

Aside from the operator \eqref{LV_dim6_Fterm}, we can 
construct gauge invariant LV operators from the (dimension two)
building blocks  
$\overline{\Phi}_\pm e^{\pm 2e V} \Phi_\pm$, $\Phi_- \Phi_+$, 
$\overline{\Phi}_- \overline{\Phi}_+$, $D_\alpha W_\beta$ and
$\overline{D}_{\dot\alpha} \overline{W}_{\dot\beta}$ with possible
gauge covariant derivatives inserted. From the identity 
\(
[ \nabla_\mu, \nabla_\nu] \Phi_\pm = \pm e \, 
(\epsilon^T \sigma)^{\alpha\beta} \nabla_\alpha( W_\beta \Phi_\pm)
\) 
and the anti-chirality of $\overline{\Phi}_\pm$, we infer that 
\begin{equation}
\int d^4\theta\, \overline{\Phi}_\pm e^{\pm 2eV} 
[ \nabla_\mu, \nabla_\nu] \Phi_\pm  
~=~ \int d^4\theta\, \Phi_- [ \nabla_\mu, \nabla_\nu] \Phi_+ ~=~ 0~. 
\end{equation} 
Moreover, full superspace integrals of 
$\Phi_- \Phi_+\, D_\alpha W_\beta$, 
$\Phi_- \Phi_+\, \overline{D}_{\dot\alpha} \overline{W}_{\dot\beta}$ 
and their conjugates vanish as well. Therefore, the most general
(genuine K\"ahler and non-reducible to superpotential) dimension six
LV matter Lagrangian is given by  
\begin{gather}
{\cal L}_{\rm LV\ dim\ 6}^{\rm matter}  ~=~ \frac{1}{M^2}
\int d^4 \theta\, \Big[ 
\overline{\Phi}_\pm e^{\pm 2eV} \Phi_\pm \, 
\Big( 
A_\pm^{\mu\nu} \, D \sigma_{\mu\nu} W ~+~ 
\overline{A}_\pm^{\mu\nu} \, \overline{D} \bar\sigma_{\mu\nu} \overline{W}
\Big) 
\nonumber \\[2ex]
~+~ S_\pm^{\mu\nu}\,  \overline{\Phi}_\pm e^{\pm 2eV} 
\lbrace \nabla_\mu, \nabla_\nu \rbrace \Phi_\pm  
~+~ Z^{\mu\nu}\,  \Phi_- \lbrace \nabla_\mu, \nabla_\nu \rbrace \Phi_+ 
~+~ \overline{Z}^{\mu\nu}\,  
\overline{\Phi}_- \lbrace \nabla_\mu, \nabla_\nu \rbrace \overline{\Phi}_+ 
 \Big]~, 
 \label{LV_dim6_Dterm}
\end{gather}
where $S_\pm^{\mu\nu}$ are real symmetric traceless matrices, 
$Z^{\mu\nu}$ is a complex symmetric traceless matrix and
$\overline{Z}^{\mu\nu}$ is its complex conjugate.

In this section, we do not give the explicit component expressions of
these supersymmetric operators, but it is not hard to see that
operators like  $ F_{\mu\rho}F_{\nu\sigma}F^{\rho\sigma} $
or $ F_{\rho\sigma}F^{\rho\sigma}F_{\mu\nu} $, 
do not arise. This might seem surprising, since such terms do appear
in investigations of non-commutative SUSY models, and SQED in
particular \cite{Putz:2002ib,Mikulovic:2003sq}. However, there is no
inconsistency: as pointed out in \cite{Mikulovic:2003sq} the
Seiberg-Witten map for non-commutative supersymmetric gauge 
theories cannot simultaneously have local and chiral gauge
transformations and be invariant under conventional supersymmetry. 
In our construction we have insisted on these three
principles. 
Thus our framework is more restrictive and does not allow 
for the operators cubic in the electromagnetic field strength.

\section{Quantum corrections in the presence of LV}
\label{quantum}

\subsection{Absence of a LV induced D-term}
\label{noDterm}

In this subsection we want to show that the dimension five LV operators
discussed in section \ref{Dim5LV} do not lead to dangerous power 
law divergences in SQED. Before we enter this analysis, we would like
to emphasize why this is an important issue. 

One of the main reasons supersymmetry is conventionally introduced is
that supersymmetric theories are free of destabilizing quadratic
divergences. There is of course one well-know 
exception to this assertion, the $D$-term of a U(1) vector multiplet,
which is in principle quadratically divergent at one loop. 
However, in any supersymmetric theory that is free of anomalies the
coefficient in front of the $D$ term vanishes identically. The
introduction of higher dimensional LV operators could upset the fine
balance of the cancellation of the $D$-term, reintroducing a quadratic
divergence. We will now show that such destabilizing effects do not
arise.

To begin this investigation, we specify the relevant Feynman rules. 
The matter LV operators \eqref{LV_matter} can be decomposed into
a modification of the kinetic term of the chiral multiplets 
\begin{equation} 
\int d^4\theta\, 
\overline{\Phi}_\pm 
\Big( 1 ~+~ \frac{N_\pm^\mu}{M} \, i \partial_\mu  \Big) 
\Phi_\pm~, 
\label{LV_matter_int00}
\end{equation}
and their gauge interactions:
\begin{equation} 
\raisebox{-1ex}{\includegraphics[height=1cm,keepaspectratio]{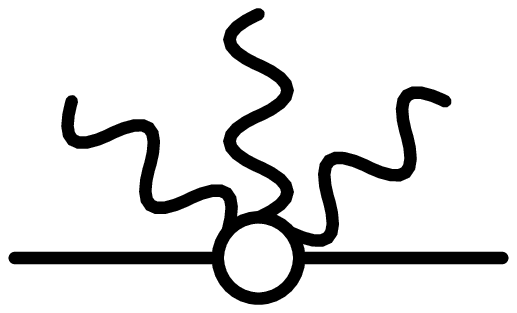}}
\quad = \quad 
\int d^4\theta\, 
\overline{\Phi}_\pm 
\Big(  e^{\pm 2 e V} ~-~ 1 \Big) 
\Big( 1 ~+~ \frac{N_\pm^\mu}{M} \, i \partial_\mu  \Big) 
\Phi_\pm~, 
\label{LV_matter_int1}
\end{equation}
and 
\begin{equation}
\raisebox{-1ex}{\includegraphics[height=1
cm,keepaspectratio]{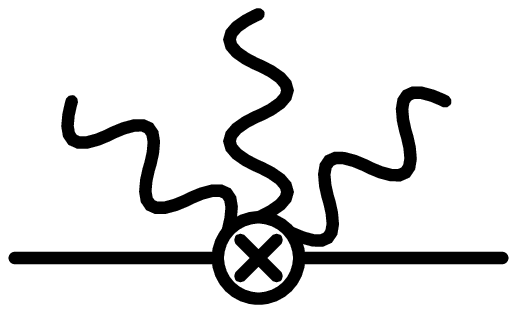}}
\quad = \quad 
\pm \int d^4\theta\, 
\frac {e \, N^\mu \, \bar\sigma_\mu^{\dot\alpha\alpha}}{2M} ~ 
\overline{\Phi}_\pm 
e^{\pm 2eV} 
( \overline{D}_{\dot\alpha} D_\alpha V) \, 
\Phi_\pm~. 
\label{LV_matter_int2}
\end{equation} 
For most  phenomenological applications it is sufficient to include
only the first order terms in expansion in LV parameters. For 
the study of the $D$-term, however, higher order terms in LV have to be
taken into account as well. It proves useful to combine quadratic
terms (\ref{LV_matter_int00}) into re-summed  propagators 
\begin{equation} 
\raisebox{-1ex}{\includegraphics[width=1.4cm,keepaspectratio]{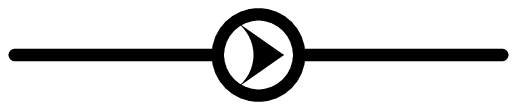}}
\quad = \quad 
\frac 1{\Box}~  \frac 1{1 ~+~  \frac{ N_\pm^\mu} M \, i\partial_\mu}~. 
\end{equation} 
These propagators are better behaved in the UV, because of the additional 
derivative in the denominator. Since the momentum scale involved in
the $D$-term calculation is far above the soft breaking scale, we
ignore soft scalar masses and the electron mass. The LV parts of the
re-summed propagators are canceled exactly by the corresponding parts of
the interactions \eqref{LV_matter_int1}, when these propagators are
attached to their $\Phi_\pm$-legs. Diagrammatically this may be 
represented as  
\begin{equation}
\raisebox{-1ex}{\includegraphics[height=1cm,keepaspectratio]{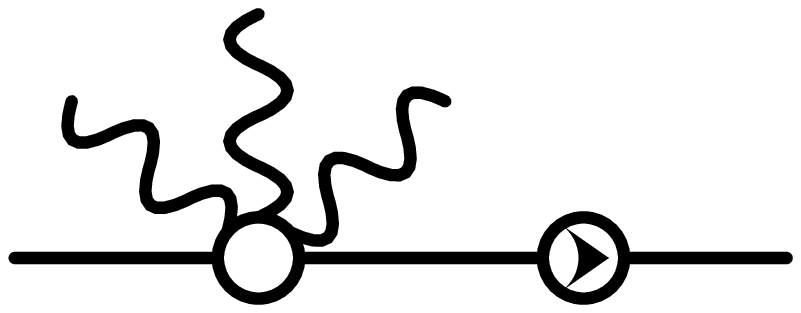}}
\quad = \quad 
\raisebox{-1ex}{\includegraphics[height=1.1cm,keepaspectratio]{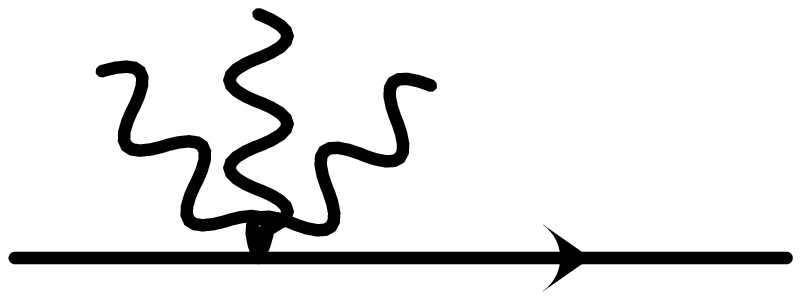}}
~. 
\label{ResummedPropVer}
\end{equation}
This shows that for a single insertion of the interactions
(\ref{LV_matter_int1}) and \eqref{LV_matter_int2} only the latter
survives, leading to logarithmic renormalization of the dimension five
LV operators, which will be studied in the next subsection. 
Another immediate consequence of (\ref{ResummedPropVer}) is that LV
does not  modify the cancellation of the Fayet-Iliopoulos (FI)
$D$-term at one loop. Indeed, the 
$\overline{D}_{\dot\alpha}D_\alpha V$-
proportional interaction gives a total derivative in the superspace
when $\Phi_{\pm}$ fields are integrated out, and thus vanishes. 
The part of interaction  \eqref{LV_matter_int1} linear in $V$ could
induce the $D$-term via the tadpole diagrams obtained by closing the
chiral loop in the diagrams above. However, cancellation  property
(\ref{ResummedPropVer}) reduces the tadpole with LV to a standard 
tadpole diagram of the Lorentz-preserving case, 
\begin{equation}
\raisebox{-2.5ex}{\includegraphics[height=1.2cm,keepaspectratio]{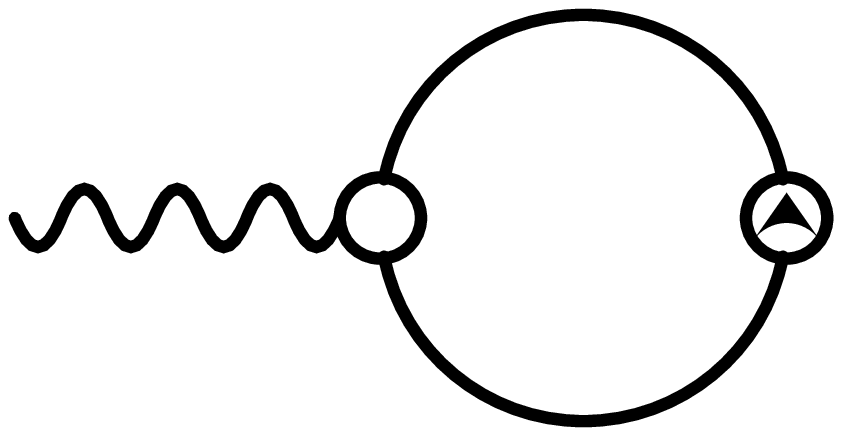}}
\quad = \quad 
\raisebox{-2.5ex}{\includegraphics[height=1.2cm,keepaspectratio]{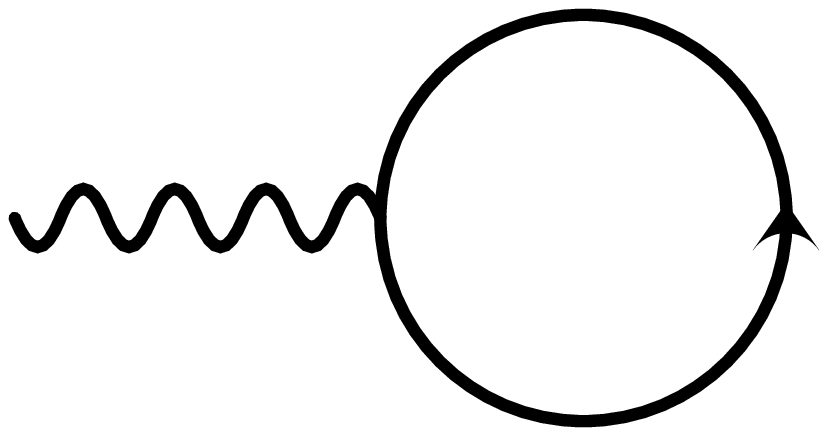}}
\quad = \quad 0~,
\end{equation}
where the last diagram gives a vanishing $D$-term when both $\Phi_+$
and $\Phi_-$ loops are taken into account. More generically, the
$D$-term will vanish for any chiral field content, provided that the
sum of all charges of chiral fields is zero. Hence, to first order in
the LV parameters, no extra quadratic divergences are introduced into
SQED by LV interactions.

The situation becomes more complicated if we go to higher orders in 
the LV parameters and to higher loop orders: the arguments 
presented above are sufficient to
prove that to all orders no quadratic divergences arise, as long as we
ignore the second interaction structure \eqref{LV_matter_int2}. At two-loop 
level vertex \eqref{LV_matter_int2} introduces additional factors of 
$\overline{D}_{\dot\alpha} D_\alpha$ into diagrams, and thereby raises 
the degree of divergence of a diagram by one, as 
\(
\{\overline{D}_{\dot\alpha}, D_\alpha\} = 
-2 i \sigma_{\alpha\dot\alpha}^\mu \partial_\mu~.
\) 
Unlike in the one-loop calculation of the FI tadpole, the 
$\overline{D}_{\dot\alpha} D_\alpha$ derivatives may now act inside
the diagrams, and hence still can lead to a potential power-like
divergence. In addition, each of the internal propagator lines may be
dressed with multiple LV insertions.

Even though the cancellation
property we relied upon at one loop, Eq.\ (\ref{ResummedPropVer}), does
not apply here,  luckily, one can show that at two (and higher) loops
the effects of all such possible insertions still cancel. The proof of
this statement is similar to the proof in a standard Lorentz
preserving U(1) theory \cite{Fischler:1981zk}. At two loops, there are
two types of diagrams, 
\begin{equation}
\raisebox{-3ex}{\includegraphics[height=1.4cm,keepaspectratio]{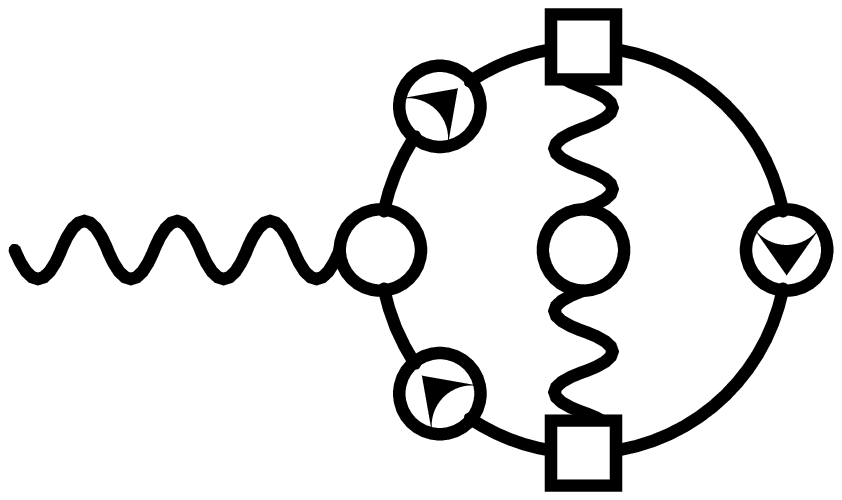}}
\quad + \quad 
\raisebox{-3ex}{\includegraphics[height=1.4cm,keepaspectratio]{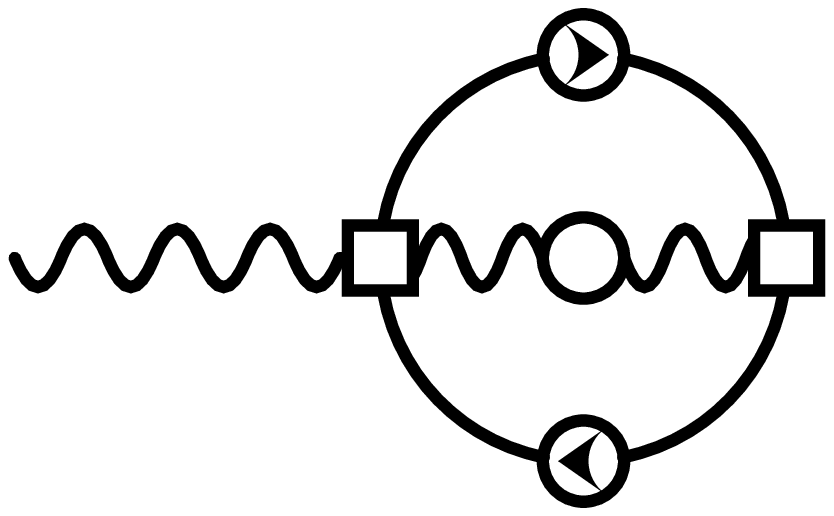}}
~,
\end{equation} 
where the vertices with boxes denote either regular gauge interactions or
the ones given in \eqref{LV_matter_int2} with the derivatives
$\overline{D}_{\dot\alpha} D_\alpha$ acting on the internal gauge lines. 
Using the diagrammatic result \eqref{ResummedPropVer}, the vertex of 
the first diagram with the external $V$ line and one adjacent chiral
line can be turned into an ordinary Lorentz-preserving combination. 
By partial integration on
the internal gauge line all (LV) operators can be moved away as far as
possible from the vertex with the external gauge multiplet. After
these manipulations the diagrams can be represented pictorially as  
\begin{equation}
\raisebox{-3ex}{\includegraphics[height=1.4cm,keepaspectratio]{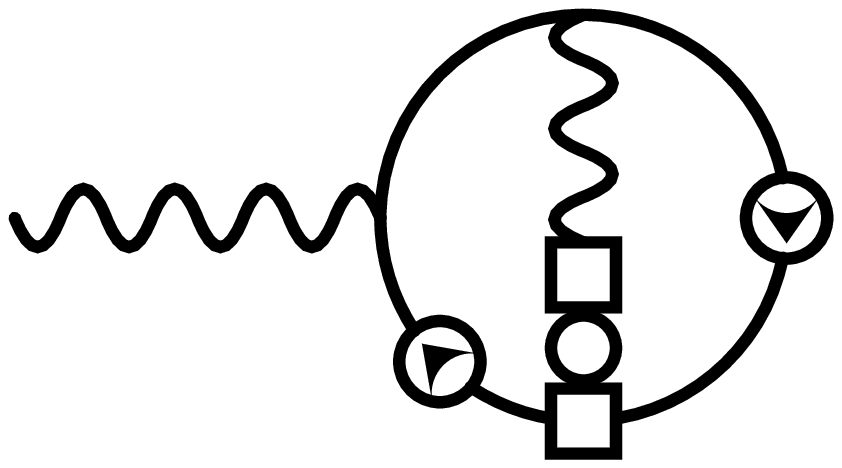}}
\quad + \quad 
\raisebox{-3ex}{\includegraphics[height=1.4cm,keepaspectratio]{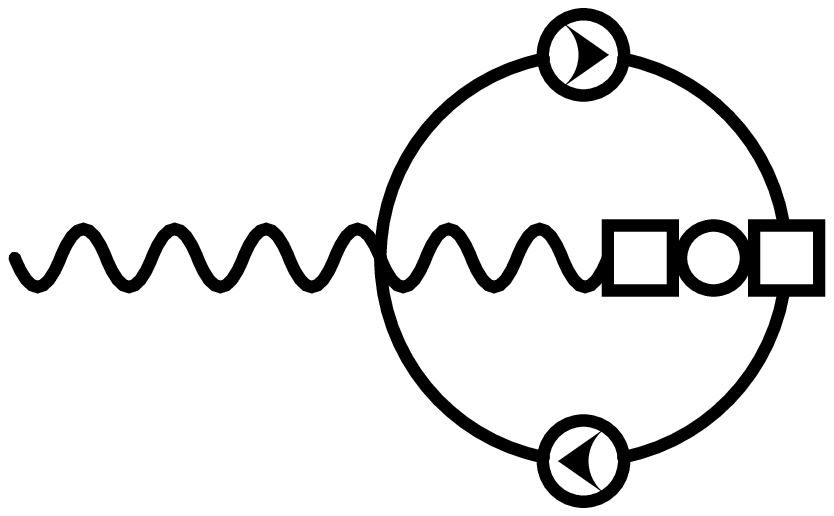}}
~.
\label{eq26}
\end{equation} 
After some straightforward algebra involving $\overline{D}{}^2$ and 
$D^2$ along the chiral field propagators, one can show that 
the ordinary chiral line in the first diagram of (\ref{eq26}) 
can be reduced to a delta function in the superspace.
This makes both diagrams in (\ref{eq26}) identical in structure, but
with opposite signs. Thus, we observe that these diagrams indeed
cancel, and no FI $D$-term arises even at two (or higher) loop level.

\subsection{Absence of LV induced gauge anomalies and of the Chern-Simons term}
\label{noAnomaly}

It is well known that anomalies put severe restrictions on the matter
spectrum of particle physics models. One may wonder whether LV 
might lead to new anomalies. Should this happen, either the LV vectors
must be restricted by stringent conditions that ensure the 
anomaly cancellation, or gauge non-invariant terms would have to be
included in the classical action in order to cancel the gauge
variation of the effective action obtained by integrating out the
fermions. In the SUSY LV context the supersymmetric extension of the
CS term would be a possible term that could cancel new  
anomalies. We will show now that LV terms at dimension five
do not modify the chiral anomaly. As a consequence, there are no
further restrictions on the LV vectors and the 
local gauge non-invariant SUSY CS term \eqref{locCS} is not
admissible. In addition to this indirect anomaly argument against the
CS term, we show  that it is not
generated by explicitly computing relevant diagrams.

To prove the claim that there are no new gauge anomalies, we 
closely follow the computation of the covariant anomaly presented in 
refs.\ \cite{Hayashi:1998ca,Gates:2000gu} using the techniques
developed by Fujikawa and Konishi
\cite{Fujikawa:1983bg,Konishi:1985tu}. We consider the classical LV
chiral multiplet action  
\begin{equation}
S ~=~ \int d^8z\, \overline{\Phi}_+ 
e^{2eV} \Big(1 ~+~ i N_+^\mu\, \nabla_\mu  \Big) \Phi_+~. 
\end{equation}
The variation of the effective action, obtained by integrating out the
chiral multiplet $\Phi_+$, under a chiral gauge transformation 
($\delta \Lambda \neq 0, \delta \overline{\Lambda} = 0$),
is given by 
\begin{equation}
\delta_\Lambda\Gamma(V) ~=~
\langle \delta_\Lambda S \rangle ~=~  2e\, 
\Big\langle \int d^8z\,  \overline{\Phi}_+ 
e^{2eV} \Big(1 ~+~ i N^\mu_+ \nabla_\mu  \Big) (\delta \Lambda\, \Phi_+)
\Big\rangle~.
\label{vareffact}
\end{equation} 
This expression is regularized by inserting the operator 
$\exp (\Box_+/M^2)$, where $\Box_+$ is the covariant d'Alembertian that
preserves chirality, and $M$ is a regulator mass which will be taken
$M \rightarrow \infty$ at the end of the computation. To
evaluate the regularized amplitude we determine the propagator in the
background field $V$: 
\begin{equation}
\langle \Phi_{+2} \overline{\Phi}_{+1} e^{2eV}_1 \rangle ~=~ 
i\,  \Big(   
1 ~+~ i N^\mu_+ \frac 1{16 \Box_+}\,  \overline{\nabla}{}^2 \nabla{}^2
\nabla_\mu
\Big)^{-1}_2 \, 
\Big( 
\frac 1{16 \Box_+} \, \overline{\nabla}{}^2 \nabla{}^2
\Big)_2 \, \delta^8_{21}~.
\end{equation}
Here the subscripts $1$ and $2$ indicate that the corresponding
expression is a function (or derivative) of the superspace coordinates
$z_1$ and $z_2$. By inserting this propagator in the variation of the
effective action \eqref{vareffact}, one can show that the LV
factors exactly cancel out, and the anomaly reduces to the standard
one without any LV. Notice that in this derivation we
have not used any properties of the operator $i N^\mu_+ \nabla_\mu$
except that it is gauge covariant. Therefore, this argument 
shows that no kinetic modification ever leads to new anomaly
constraints.

\begin{figure}
\begin{center}
\begin{tabular}{ccccc}
 \includegraphics[width=2.7cm,height=2.7cm,keepaspectratio]
 {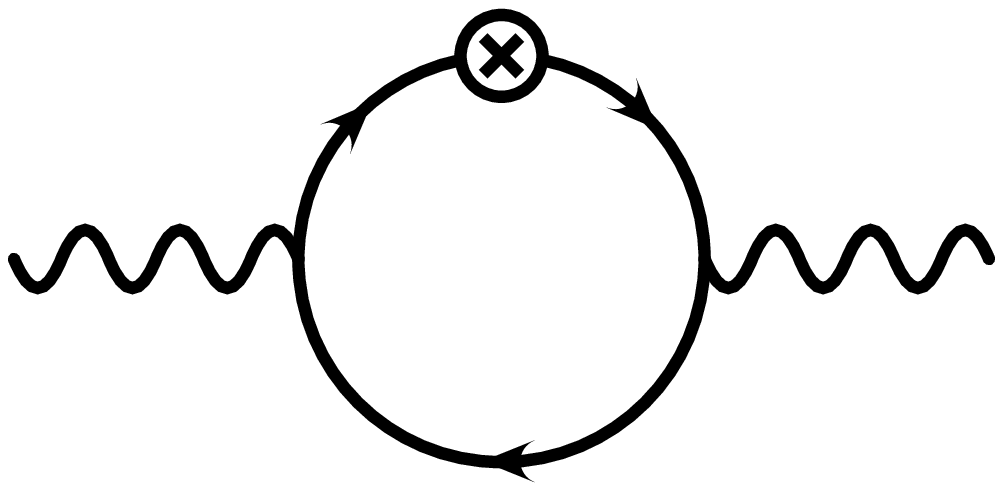} &
 \includegraphics[width=2.7cm,height=2.7cm,keepaspectratio]
 {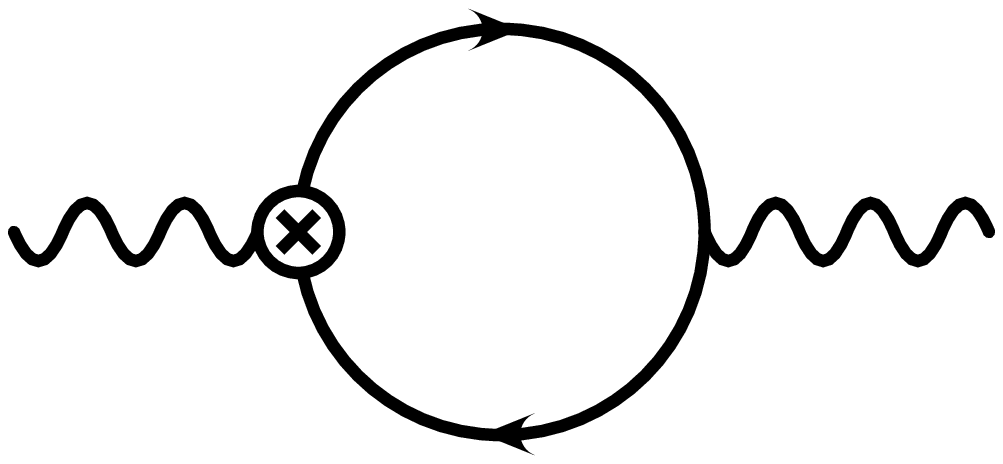} 
 \includegraphics[width=2.7cm,height=2.7cm,keepaspectratio]
 {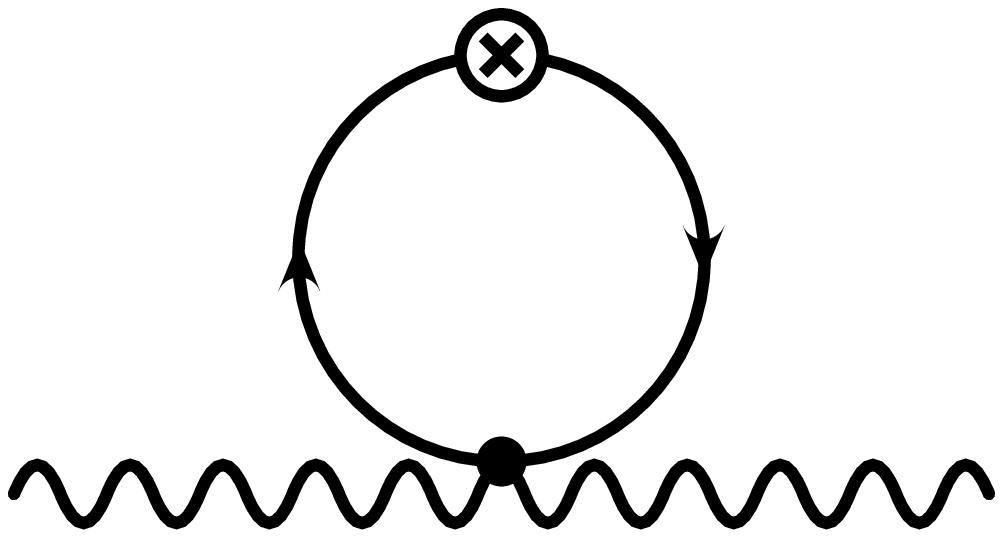} &
 \includegraphics[width=2.7cm,height=2.7cm,keepaspectratio]
 {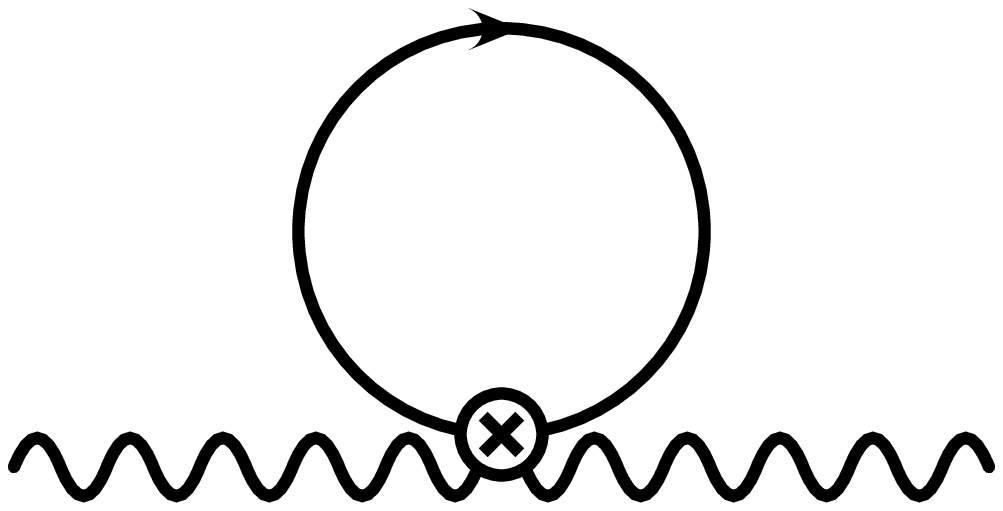} &
 \includegraphics[width=2.7cm,height=2.7cm,keepaspectratio]
 {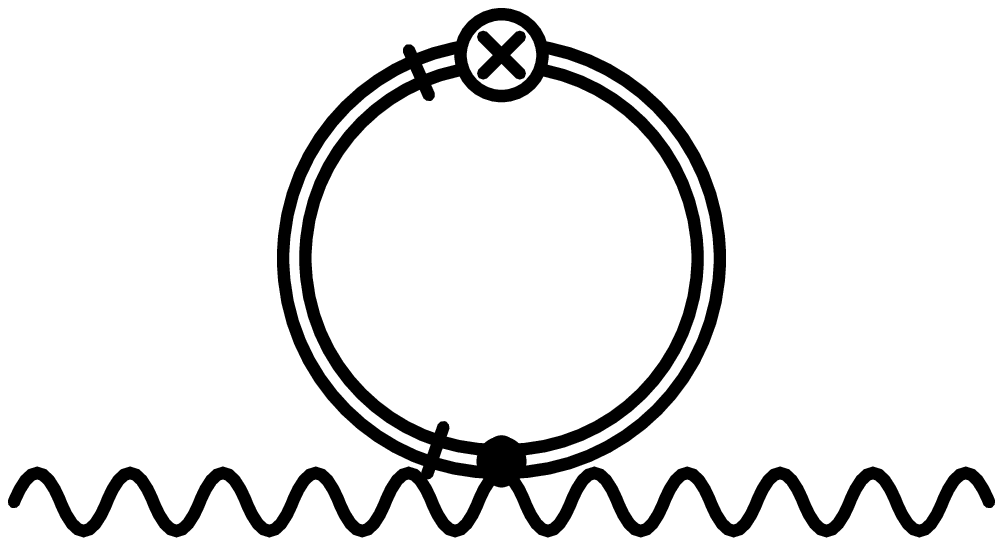} 
\end{tabular} 
\\[2ex]
\begin{tabular}{cccc}
 \includegraphics[width=2.7cm,height=2.7cm,keepaspectratio]
 {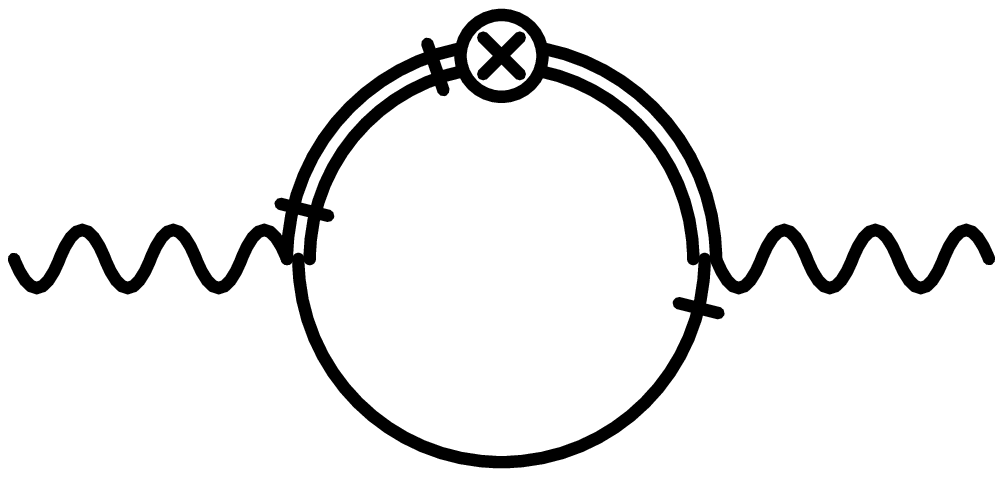} 
\raisebox{12mm}{\includegraphics[width=2.7cm,height=2.7cm,angle=180,keepaspectratio]
 {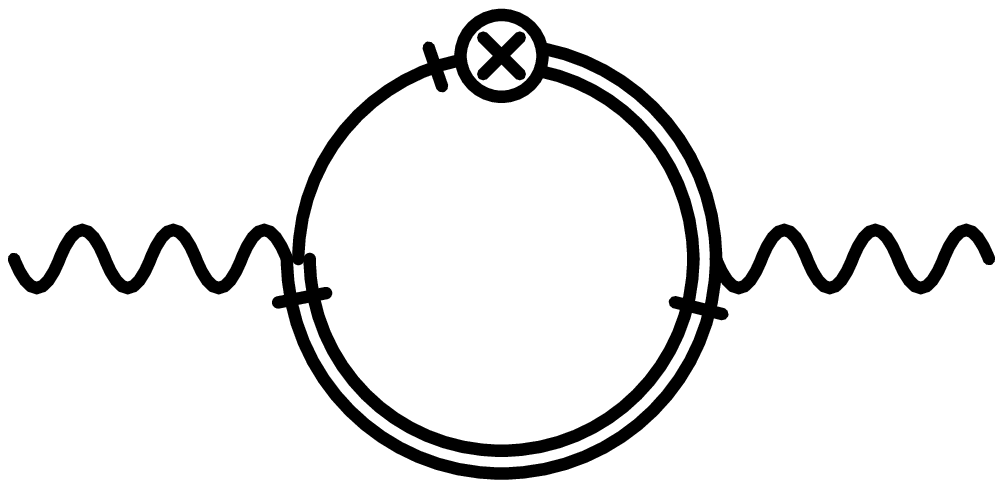}} &
 \includegraphics[width=2.7cm,height=2.7cm,keepaspectratio]
 {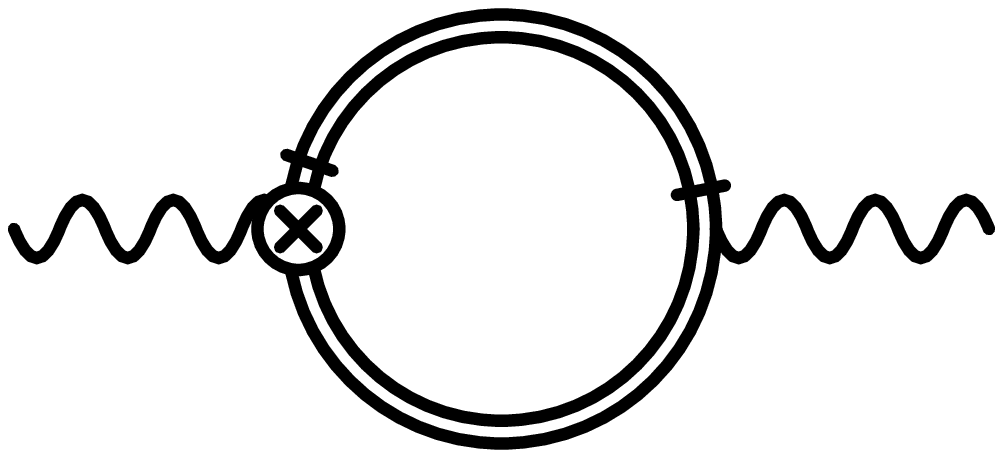} &
 \includegraphics[width=2.7cm,height=2.7cm,keepaspectratio]
 {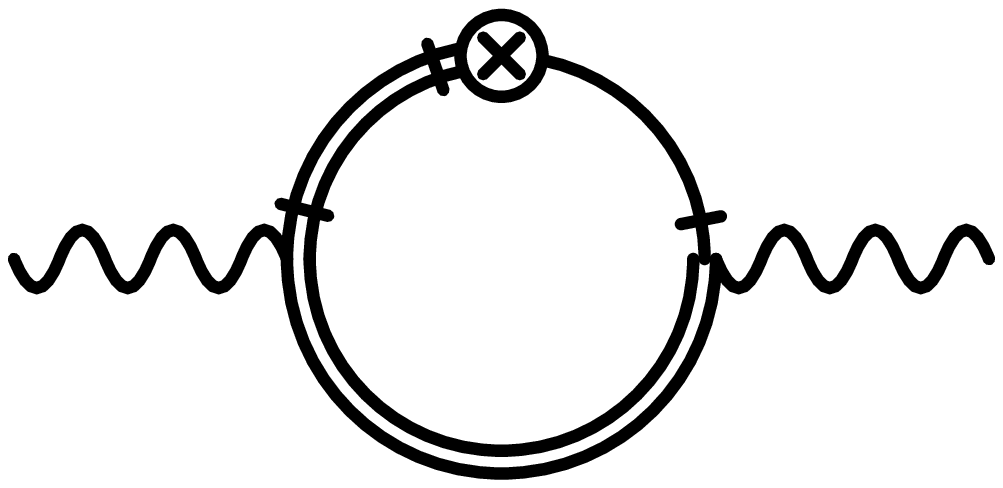} 
\end{tabular}
\end{center}
 \caption{\label{diag_gauge_massive}
  LV diagrams in massive SQED. Solid lines denote chiral field
propagators, wiggled line represent external gauge superfield
legs. Crossed circles indicate insertions of the  LV operator
(\ref{LV_matter}). Double lines represent chirality-flipping
propagators $ \langle \Phi \Phi \rangle $ and $ \langle
\overline{\Phi} \, \overline{\Phi} \rangle $. Bars denote the $
\overline{\Phi} $ ends of the propagators. Only the $ N_+^\mu $
operator is included in this figure; the $ N_-^\mu $ operator
generates the same set of diagrams. 
}
\end{figure}

Because the gauge anomaly is the same as in the Lorentz preserving
theory (and therefore absent in LV SQED), we conclude that no CS term
can be generated by quantum effects. The reason is that the local
version of the SUSY CS \eqref{locCS} is not gauge invariant, see
\eqref{locCStrans}. The gauge invariant version of the SUSY CS
\eqref{nonlocCS} is, in its turn, nonlocal. Since only local and
gauge-invariant counterterms can arise in a non-anomalous quantum
field theory, neither version of super-CS can get induced. 
This result can be confirmed by a direct loop computation: the 
diagrams of Fig.~\ref{diag_gauge_massive} can potentially generate 
the SUSY CS term \eqref{locCS}, but an explicit calculation reveals
that all these contributions cancel for both $N^\mu_+$ and $N^\mu_-$
backgrounds, even when the electron mass $m_e$ is retained. 
The absence of CS term induced by quantum effects \eqref{locCS} 
can be understood as a SUSY version of the no-go theorem of  
Coleman and Glashow \cite{CG}. In section \ref{SB_gauge_sector} we 
show that the CS term is also not induced by the soft supersymmetry
breaking.

\subsection{RGE evolution of dimension five LV operators in SQED}
\label{RGEvolution}

As mentioned in the Introduction, within an effective field theory
approach, we are allowed to assume that operators (\ref{LV_matter}), 
(\ref{LV_gauge}) and (\ref{LV_gauge_Tterm}) are generated at the UV
scale $M$ by some unspecified LV dynamics.  
All experimental limits are obtained at much lower energy
scales. Therefore, in order to derive meaningful experimental
constraints on parameters of LV SQED, we have to evolve the LV
operators down to the low-energy scale. Furthermore, we know that
SUSY is broken, and the operators of dimension five will
source dimension three LV operators via SUSY breaking, leading to
tight bounds on LV parameters of the model. In this section, we derive
and solve the renormalization group equations (RGE's) for dimension
five LV operators assuming unbroken SUSY. In the next section we
include the effects of soft breaking and calculate resulting dimension
three operators.

We work in the linear approximation in LV parameters, and
neglect all terms that involve higher powers of $1/M$.
The running of the LV operators (\ref{LV_matter}), (\ref{LV_gauge})
and (\ref{LV_gauge_Tterm}) is, in part, a consequence of the wave
function renormalization of various superfields induced by
standard SQED one-loop diagrams. We do not give them explicitly here,
but we take their effects into account in the resulting RGE's. For the
logarithmic running of the LV parameters above the supersymmetric
threshold, we can ignore soft breaking masses and electron mass inside
loops. At one loop, this means that loop diagrams with internal
lines of $\Phi_+$ and $\Phi_-$ can be calculated independently.

\begin{figure}
\begin{center}
\begin{tabular}{cccc}
\includegraphics[width=2.7cm,height=2.7cm,keepaspectratio]{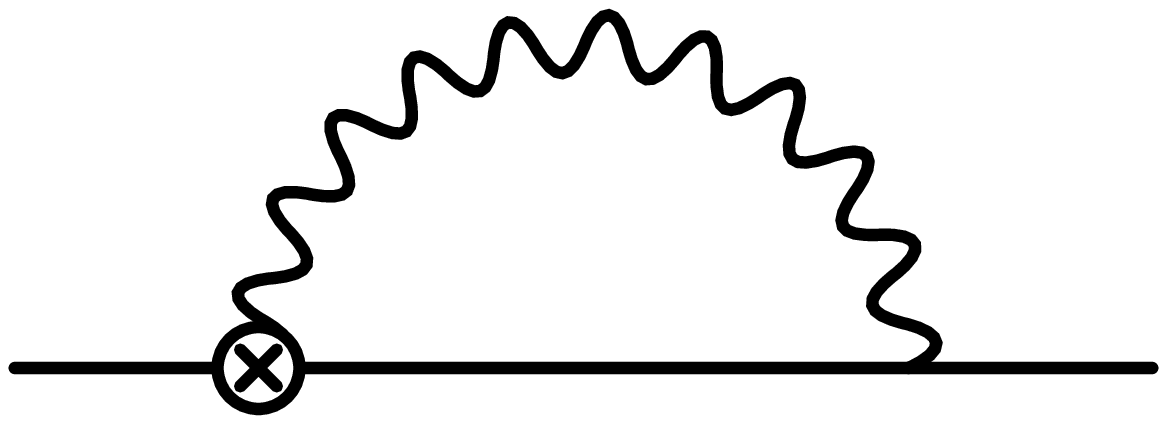}
&
\includegraphics[width=2.7cm,height=2.7cm,keepaspectratio]{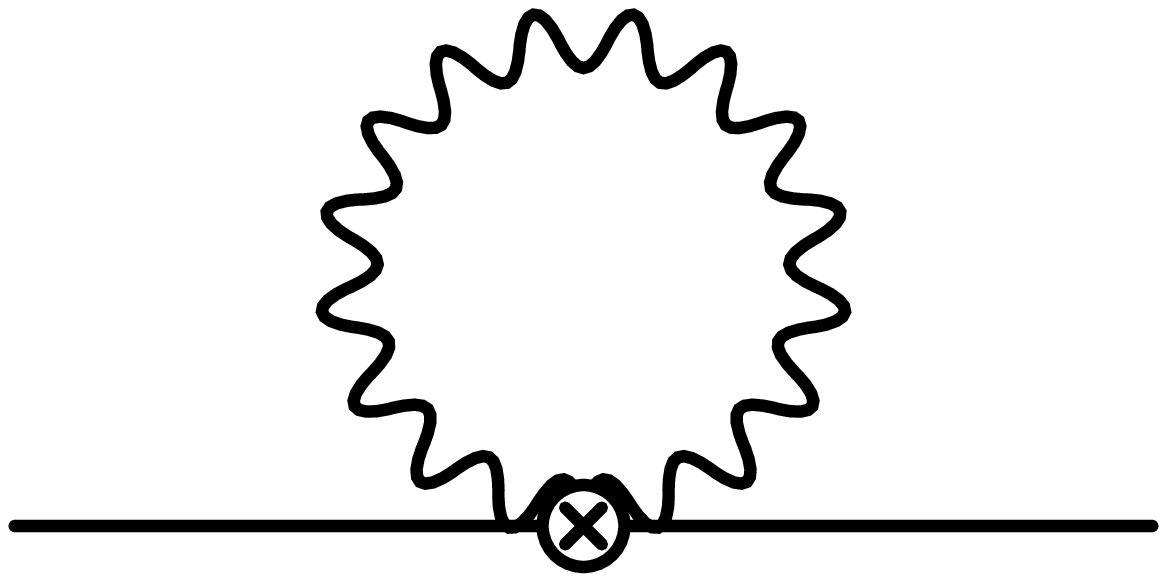}
&
\includegraphics[width=2.7cm,height=2.7cm,keepaspectratio]{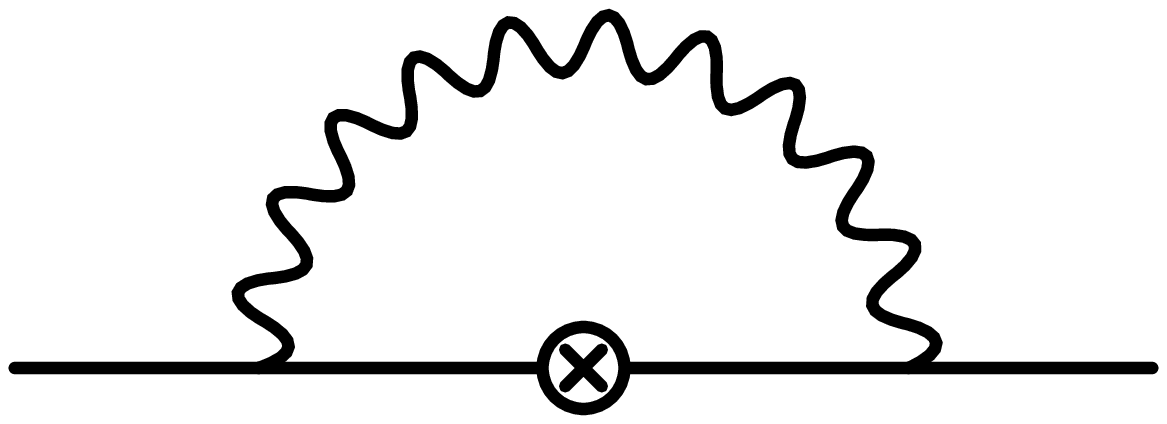}
&
\includegraphics[width=2.7cm,height=2.7cm,keepaspectratio]{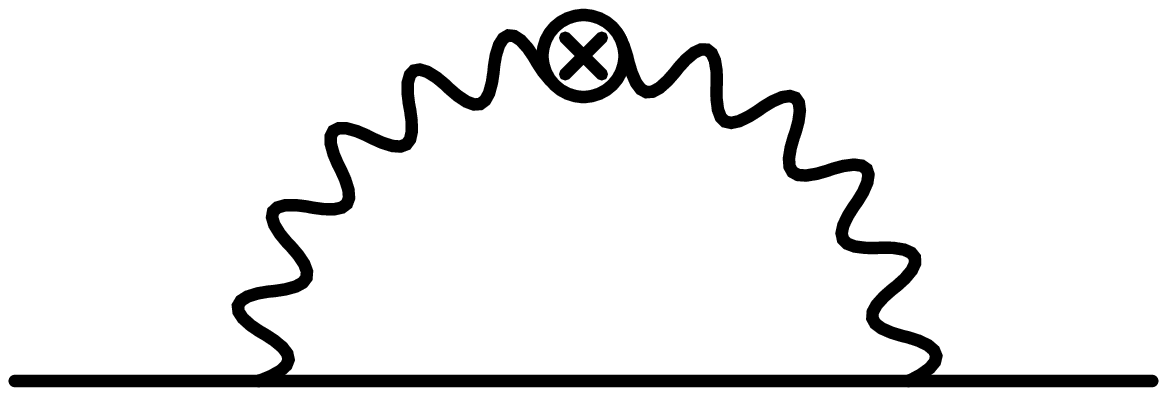}
\end{tabular}
\end{center}
\caption{\label{diag_LV_chiral}
One-loop corrections to the matter multiplet operators
(\ref{LV_matter}). 
The notations are the same as in Fig.~\ref{diag_gauge_massive}.
Wiggled lines represent the gauge superfield propagators, and the
crossed circles are insertions of the LV operators (\ref{LV_matter})
or (\ref{LV_gauge}). 
}
\end{figure}

\begin{figure}
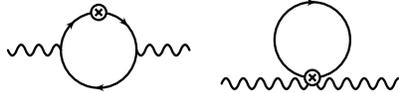

\begin{center}
\begin{tabular}{cc}
\includegraphics[width=2.7cm,height=2.7cm,keepaspectratio]{diag_gauge_A.ps}
&
\includegraphics[width=2.7cm,height=2.7cm,keepaspectratio]{diag_gauge_F.ps}
\end{tabular}
\end{center}
\caption{\label{diag_LV_gauge}
One-loop corrections to the gauge LV operator 
\eqref{LV_gauge} using the same pictorial notation as in 
Fig.~\ref{diag_LV_chiral}. 
}
\end{figure}

The renormalization of the electron/positron LV operators
\eqref{LV_matter} is induced by the diagrams shown in 
Fig.~\ref{diag_LV_chiral}. The first two diagrams involve the
interactions \eqref{LV_matter_int2}. Notice that the seagull diagram
vanishes because the photon superfield loop contains only two super
covariant derivatives. The last diagram is induced by the operator  
\eqref{LV_gauge}. Loops with a single insertion of tensor
interaction \eqref{LV_gauge_Tterm} vanish identically, as there are no
operators in the chiral sector that can couple to $T^{\mu\,\nu\lambda}$.

In the gauge sector we find that the renormalization of the tensor LV
gauge operator \eqref{LV_gauge_Tterm} is absent. Indeed, since we 
work in the first order in LV, this operator cannot receive any
corrections from operators that depend on vector backgrounds. The
renormalization of LV gauge operator, with $N^\mu$ \eqref{LV_gauge}, 
is given by the diagrams shown in Fig.~\ref{diag_LV_gauge}. 
Again, one can use the cancellation property (\ref{ResummedPropVer}) 
to observe that the vertex in the second diagram of
Fig.~\ref{diag_LV_gauge} is given by \eqref{LV_matter_int2} only. The 
combination of these gauge self-energy diagrams is only
logarithmically divergent, which gives another reason why the 
dimension three LV CS term is not generated by loop effects
in this approximation.

After a straightforward calculation of logarithmically divergent parts
of the diagrams in Figs.~\ref{diag_LV_chiral} and \ref{diag_LV_gauge},
and inclusion of wave function renormalization effects, we arrive at
the renormalization group equation (RGE) for the LV parameters: 
\begin{equation}
\label{RG_eqn_undiag}
     \mu \frac{\partial}
              {\partial\mu} 
                \left(
\begin{array}{c}
                   N^\nu \\ 
   N_+^\nu \\
                   N_{-}^\nu \\
   T^{\mu\,\nu\rho}
                \end{array} \right) ~=~  
     \frac{\alpha}
          {2 \pi} \, 
     \left(\begin{array}{rrrr}
                    2 & -1 & -1 & ~~0 \\
   -6 &  3 &  0 & ~~0 \\
                   -6 &  0 &  3 & ~~0 \\
    0 &  0 &  0 & ~~2
           \end{array}\right)
     \left(
  \begin{array}{c}
                 N^\nu \\ 
 N_+^\nu \\
                 N_{-}^\nu \\
 T^{\mu\,\nu\rho}
          \end{array} \right)~,
\end{equation}
As usual, $\alpha = e^2/(4\pi)$ denotes the fine structure constant. 
The (1,1) and (4,4) elements of the matrix in (\ref{RG_eqn_undiag})
are equal and  result only from the renormalization of wave functions.
The electron and positron LV parameters $N_\pm^\mu$ both give and
receive equal contributions to and from the vector LV parameter
$N^\mu$. This explains why the pairs of matrix elements (1,2) and (1,3),
(2,2) and (3,3), and (2,1) and (3,1) are equal.  

It will prove useful to introduce the following combinations of 
LV parameters that couple to operators of definite parity:
\begin{equation}
\label{def_Nmu}
      N^\mu_V ~=~  \frac{ N_+^\mu ~-~ N_-^\mu }{2}~,   
\qquad 
      N^\mu_A ~=~  \frac{ N_+^\mu ~+~ N_-^\mu }{2}~.
\end{equation}
$N^\mu_V$ and $N^\mu_A$ are the charge conjugation odd and even
combinations, respectively. In general, vector backgrounds do not need
to have  the same orientation in Minkowski space, and the off-diagonal
elements of the renormalization group coefficients  in
(\ref{RG_eqn_undiag}) mix them, resulting in changes of their
directions.  By diagonalizing \eqref{RG_eqn_undiag} we identify a 
set of eigenvectors  
\begin{equation}
N_1^\mu ~=~ N^\mu_V~, 
\qquad 
N_2^\mu ~=~ 3 N^\mu - 2 N^\mu_A~,
\qquad 
N_3^\mu ~=~ 2 N^\mu + N^\mu_A~,
\end{equation} 
that evolve under the RGE independently; each of them may change its
size but not its direction. $N_V^\mu$ renormalizes independently
because it is the only combination that is odd under the charge
conjugation. 
In this basis, the RGE's and their solutions are given by 
\begin{equation} 
\mu \frac{\partial}{\partial\mu} \, N_i^\nu 
~=~ \lambda_i\, \frac { \alpha}{2 \pi} \, N_i^\nu~ 
\qquad \Rightarrow \qquad 
N_i^\nu(\mu) ~=~ 
\Big(  \frac {\alpha(\mu)}{\alpha(M)} \Big)^{\frac {\lambda_i}2} \, 
N_i^\nu(M)~, 
\label{LV_at_soft_scale}
\end{equation} 
where the eigenvalues read 
$\lambda_i = (\lambda_1, \lambda_2, \lambda_3) = (3, 6, -1)$. 
To obtain these solutions we have used the standard SQED beta function
\( 
\mu \frac{\partial}{\partial\mu} \, \alpha = \frac 1{\pi} \,
\alpha^2.  
\)

Within the SQED framework, the renormalization effects of these LV
parameters are small: even if we take $\mu = m_{s} \approx 1$ TeV and
$M = M_{\rm Pl} \approx 10^{19}$ GeV, the running affects the LV
parameters by only about 10\%. In other words, the linearized version of
(\ref{LV_at_soft_scale})  
\begin{equation}
N_i^\nu(m_s) ~\simeq~ 
\Big(1~-~ \frac{\lambda_i\, \alpha}{2\pi}\log (M/m_s)
\Big)\, N_i^\nu(M)
\label{simplified}
\end{equation}
gives a good approximation of the exact answer. The same conclusion
holds for the running of the irreducible tensor
$T^{\lambda\,\mu\nu}$. Although it may look as though a 10 percent
level change in $N_i^\mu$ is 
insignificant, one should keep in mind, that in a more
realistic framework of MSSM the number of charged degrees of 
freedom running inside the loops is significantly larger than in SQED, 
which would lead to appreciable changes in LV parameters 
between the Planck and the weak scales. 
Nevertheless, the main numerical
change in the actual size of observable LV effects will result from
soft SUSY breaking, as will be discussed in the next section.

\section{Induced dimension three operators by soft SUSY breaking}
\label{InducedDim3}

Once SUSY is broken, dimension three LV operators can be induced 
with coefficients controlled by the soft-breaking mass scale. 
Following the usual approach (see {\em e.g.}\ Ref.~\cite{Wess:1992cp}),
we introduce spurion superfields ($\theta^2$, $\overline\theta{}^2$) in
superspace expressions. We consider only soft SUSY breaking in the
matter sector. We ignore other soft-breaking terms, including a
gaugino mass, which can be motivated by the most common MSSM scenarios.
Generically, we can assume that parity is broken, so that
the scalar partners of left- and right-handed electrons have different masses.

The possible soft SUSY breaking masses of the electron and positron
can be written as 
\begin{equation}
\label{SB_vertex}
  \mathcal{L}_{SB} ~=~  
- \int d^2\theta ~ \theta^2~ 
(m_{s}^0)^2 \, \Phi_+ \Phi_-
~+~ \text{h.c.} 
~- \int d^4\theta ~
\theta^2\overline{\theta}{}^2~ 
\Big( 
(m_s^+)^2\, \overline{\Phi}_+ \Phi_+ 
~+~
(m_s^-)^2\, \overline{\Phi}_- \Phi_-
\Big)  
~, 
\end{equation}
where $m_s^\pm$, $m_s^{0}$ are real and complex masses, respectively. 
To make parity violation manifest in the SUSY breaking 
\eqref{SB_vertex} we introduce 
\begin{equation}
\Delta m^2 ~=~ (m_{s}^+)^2 ~ - ~ (m_{s}^-)^2~, 
\qquad
(m_{s}^\pm)^2 ~=~ m_s^2 ~\pm~ \frac{\Delta m^2}2~.
\label{deltam}
\end{equation}
The parity conserving scenario is obtained in the limit 
$ \Delta m^2 \to 0 $. Throughout the paper we assume that 
 $ \Delta m^2$ is somewhat smaller but not necessarily much smaller than 
$m_s^2$, and that the values of the soft-breaking parameters are such that 
scalar electrons do not develop vacuum expectation values. 
Viewing SQED as a subset of MSSM, we can also neglect $(m_{s}^0)^2$, $(m_{s}^0)^2
\sim O(m_s m_e)\ll m_s^2$. 

Once SUSY
is broken via (\ref{SB_vertex}) in the Lorentz-conserving sector, it
will be communicated to the LV sector via loop corrections or on the
equations of motion (EOM's), resulting in LV operators of dimension three.  

We start by listing all such operators in components, essentially
extending the existing QED parametrization \cite{Kost1}  to the SQED
field content. In the matter sector these 
operators are  
\begin{eqnarray}
{\cal L}_{\rm SB~LV~dim~3}^{\rm matter} 
&~=~& 
2\;i\, \widetilde{A}_+^\mu\, \overline{z}_+ \mathcal{D}_\mu z_+ 
~+~ 2\;i\, \widetilde{A}_-^\mu\, \overline{z}_-\mathcal{D}_\mu z_- 
~+~i\, \widetilde{C}^\mu\, z_- \mathcal{D}_\mu z_+ 
\label{LV_dim3_comp}
\\[1ex] 
\nonumber
&& 
~+~\widetilde{B}_+^\mu\, \overline{\psi}_+\overline{\sigma}_\mu \psi_+ 
~+~ \widetilde{B}_-^\mu\, \overline{\psi}_-\overline{\sigma}_\mu \psi_-
 ~+~\widetilde{D}^{\mu\nu}\, \psi_- \sigma_{\mu\nu} \psi_+~.
\end{eqnarray}
In superfield notation they can be expressed as:
\begin{gather}
{\cal L}_{\rm LV~dim~3}^{\rm matter~SB} ~=~  
\int d^4\theta~ \theta^2\overline{\theta}{}^2 \, \Big[
2i  \widetilde{A}_+^\mu\,\overline{\Phi}_+ \nabla_\mu \Phi_+
~-~ 2i \widetilde{A}_-^\mu\, \Phi_-  \nabla_\mu \overline{\Phi}_-  
~+~ \frac 12\big(\widetilde{C}^\mu  \Phi_- \nabla_\mu \Phi_+ 
~+~ \text{h.c.}\big) 
\nonumber\\
\label{LV_dim3}
~+~ \frac{1}{2} \widetilde{B}_+^\mu\, 
\overline{\nabla}\, \overline{\Phi}_+ \overline{\sigma}_\mu \nabla \Phi_+
~+~ \frac{1}{2}\widetilde{B}_+^\mu\, 
\overline{\nabla}\, \overline{\Phi}_- \overline{\sigma}_\mu \nabla \Phi_- 
~+~ \frac 12\big(
\widetilde{D}^{\mu\nu}\, \nabla \Phi_- \sigma_{\mu\nu} \nabla \Phi_+
~+~ \text{h.c.}\big)
\Big]~. 
\end{gather}
The superfield expressions (\ref{LV_dim3}) for the operators
(\ref{LV_dim3_comp}) are not unique.  One can use alternative spurion
insertions  inside gauge-invariant supersymmetric LV operators
\cite{GrootNibbelink:2004za}. However, at the component level these
expressions will reduce to linear combinations of the operators given
in (\ref{LV_dim3}).

In the gauge sector there are only two LV operators of dimension 
three: 
\begin{eqnarray}
{\cal L}_{\rm SB~LV~dim~3}^{\rm gauge} ~=~ 
\widetilde{E}_\mu\, \epsilon^{\mu\nu\rho\sigma}\, 
A_\nu \partial_\rho A_\sigma  ~+~ 
\widetilde{F}_\mu\, \lambda \sigma^\mu \overline{\lambda} 
~,
\end{eqnarray}
which can be rewritten in a superfield form using the 
CS superfield \cite{Cecotti:1987nw}:
\begin{eqnarray}
\label{CSint}
{\cal L}_{\rm LV ~dim~3}^{\rm gauge} ~=~ \int d^4\theta~ 
\Big(
(\widetilde{F}_\mu \,-\,   \widetilde{E}_\mu)\, \theta^4\, 
 W \sigma^\mu \overline{W} 
~+~ \widetilde{E}_\mu\theta \sigma^\mu\bar\theta 
\big\{ 
D^\alpha (V\, W_\alpha) 
\,+\,  
\overline{D}_{\dot\alpha}V\, \overline{W}^{\dot\alpha}
\big\}
\Big).
\end{eqnarray}

\subsection{Operators in the matter sector}

We now turn to the discussion of possible mechanisms that transmute
dimension five SUSY LV operators 
into dimension three LV operators. There are two generic ways this 
may occur, at tree level via reduction over the EOM and via loop
effects,   
\begin{equation}
\begin{array}{l c r l} 
\, [LV]_{\rm dim~5} & ~\stackrel{\mathrm {EOM}}{\longrightarrow}~ &
  (m_{s}^2 + m_e^2)\, [LV]_{\rm dim~3}~,~ &{\rm for~selectrons}~, 
\\[1ex] \,
[LV]_{\rm dim~5} &~ \stackrel{\mathrm {1\ loop}}{\longrightarrow}~ &
  m_{s}^2\, [LV]_{\rm dim~3}~,~ & {\rm for~fermions ~and ~vector~bosons}
~.
\end{array}
\nonumber
\end{equation}
The tensor operator (\ref{LV_gauge_Tterm}) does not mix with dimension
three operators in any order in SUSY breaking, because there is simply
no dimension three operator that can couple to
$T^{\mu\,\nu\lambda}$. Below we discuss in detail how dimension
three operators are generated by tree level and loop effects.

The soft supersymmetry breaking (\ref{SB_vertex}) affects LV
interactions for left- and right-handed selectrons already at tree
level. The masses of scalar particles are lifted with respect to the
masses of the electron and positron. This alters sfermions' equations
of motion, leading to an {\em enhancement} of certain dimension three
operators. Ignoring $\Delta m^2$ for a moment, one can easily show
that the combination of LV operators (\ref{LV_matter}) and SUSY
breaking (\ref{SB_vertex}) leads to the following dimension three LV
operator 
\begin{equation}
  \mathcal{L}_{\rm sparticle}^{\rm EOM} ~=~  
\frac{N_V^\mu}{M}\, 2 i\, 
\big(
m_e^2 \,+\,  m_s^2
\big)
\Bigl\{ 
\overline{z}_+ \mathcal{D}_\mu z_+ 
~-~
\overline{z}_- \mathcal{D}_\mu z_- 
\Bigr\}~,  
\end{equation}
effectively generating the $ \widetilde{A}^\mu_\pm $-terms
(\ref{LV_dim3_comp}), 
 \begin{equation}
\widetilde{A}_\pm^\mu ~=~  
\pm\, 2\, \frac{N_V^\mu}{ M }   \, 
(m_e^2 \, +\,  m_s^2)~.
\end{equation}
However, we will not be interested in these particular operators due
to the current impossibility to experimentally  study the superpartner
sector. In the matter sector only those operators involving electrons
and positrons are important for phenomenology.  
For the same reason, we ignore the possible appearance of the
operators proportional to $ \widetilde{C}^\mu $, and in the gauge
sector we will only be interested in the CS term that might
be induced for photons.

\begin{figure}
\begin{center}
\begin{tabular}{c c c}
\begin{tabular}{cc}
\includegraphics[width=2.7cm,height=2.7cm,keepaspectratio]
 {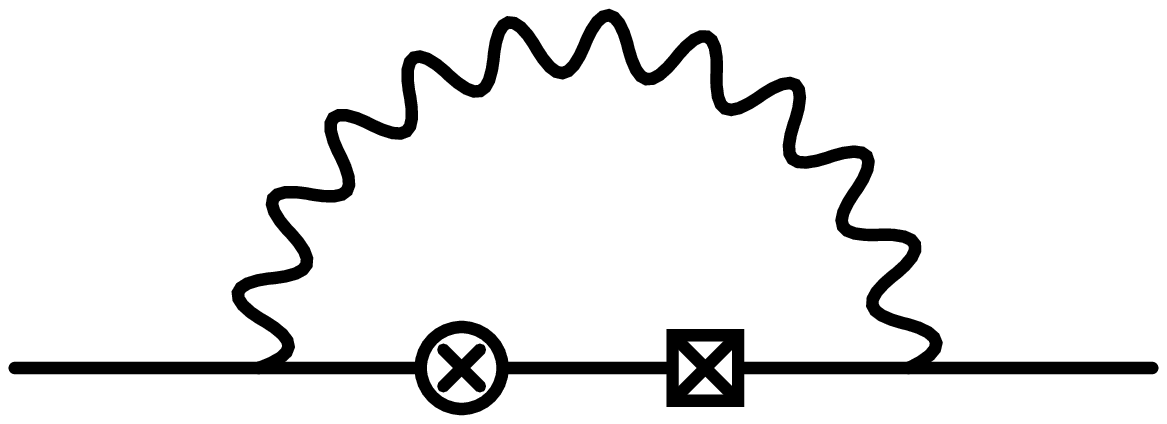} &
\includegraphics[width=2.7cm,height=2.7cm,keepaspectratio]
 {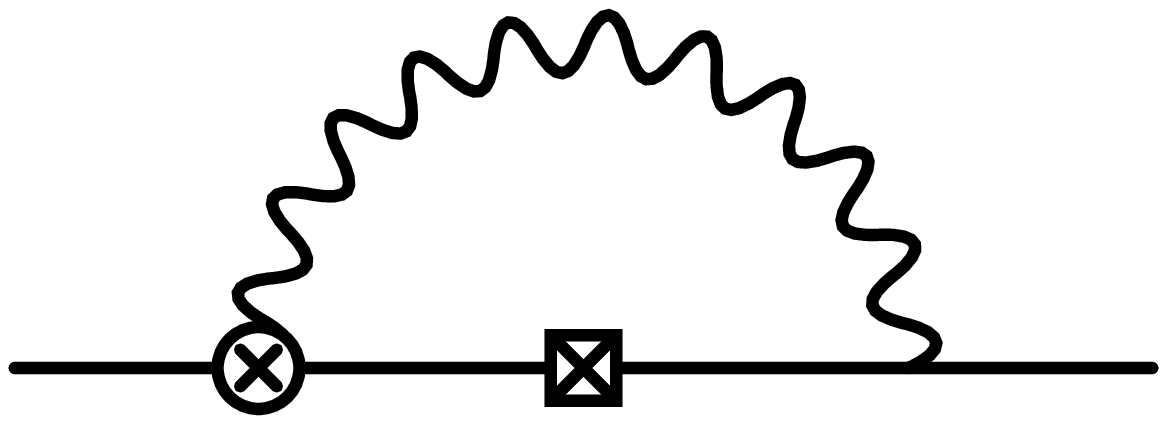} 
\end{tabular}
&\qquad\qquad \qquad  & 
\begin{tabular}{c}
\includegraphics[width=2.7cm,height=2.7cm,keepaspectratio]
 {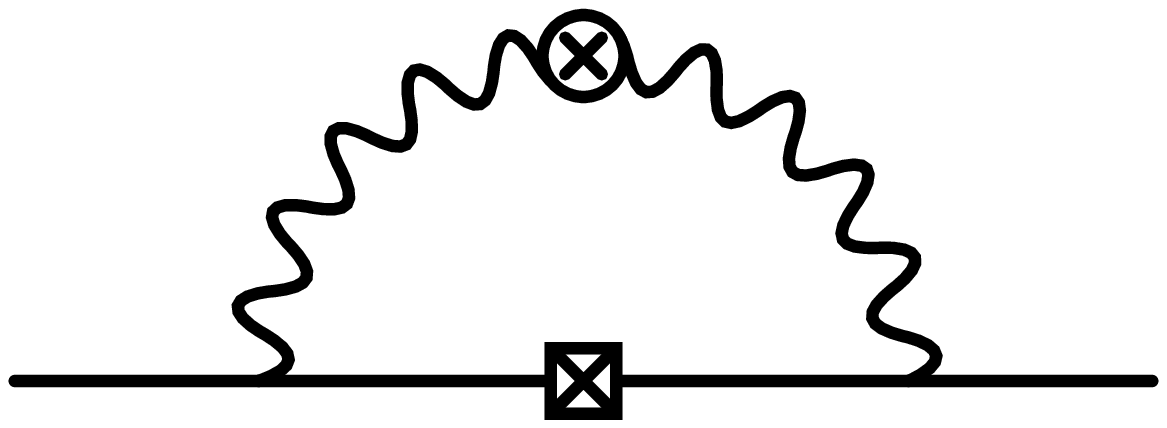}
\end{tabular} 
\\
a && b 
\end{tabular} 
\end{center}
\caption{
\label{LV_SB_chiral}
Diagrams generating dimension three LV operators for electrons
and positrons due to soft supersymmetry breaking  and dimension five
SUSY LV operators in gauge and matter sectors. In 
Figs.~\ref{LV_SB_chiral}a and \ref{LV_SB_chiral}b the inserted
operators are 
(\ref{LV_matter}) and (\ref{LV_gauge}), respectively. Finally, the box
with a cross denotes the insertion of the SUSY breaking operator
(\ref{SB_vertex}). 
}
\end{figure}

At one loop level, the transmission of SUSY breaking to the LV sector
of chiral fermions and gauge bosons may indeed be possible. We start
with one loop effects in the matter sector. It is sufficient for our
purposes to consider the running of dimension three operators  
within the interval of momenta $ m_{s} \ll |p_{\rm loop}|\ll M $, and
to retain only the contributions enhanced by a large $\log(M/m_{s})$, 
neglecting possible threshold corrections. To this accuracy, the soft
breaking parameters inside loops can be treated as perturbations 
and inserted on internal lines of diagrams from
Fig.~\ref{diag_LV_chiral}. In Figs.~\ref{LV_SB_chiral}a and
\ref{LV_SB_chiral}b we have inserted the dimension five SUSY LV
operators \eqref{LV_matter} and \eqref{LV_gauge}, respectively.

Our one loop RGE analysis concentrates on the induced dimension three 
operators $ \widetilde{B}_\mu $. Besides the contributions from
diagrams in Figs.~\ref{LV_SB_chiral}a and \ref{LV_SB_chiral}b ,the 
complete set of RGE includes one loop running of the operator $
\widetilde{B}_\mu $ itself, and its mixing with $ \widetilde{A}_\mu $.
The relevant set of RGE's includes: 
\begin{eqnarray}
\label{RG_AB}
\nonumber
\mu\frac{d \widetilde{A}_+^\nu}{d\mu} 
& ~=~ &
\frac{\alpha}{\pi} \,  \Big(  \widetilde{A}_+^\nu  
~-~\widetilde{B}_+^\nu \Big)~, 
        \\
\mu\frac{d \widetilde{B}_+^\nu}{d\mu} 
& ~=~ &
\frac{\alpha}{2\pi} \, \Big(
\widetilde{B}_+^\nu  ~-~ \widetilde{A}_+^\nu  
~+~3\, \frac{(m_s^+)^2}{M}\, N^\nu 
~-~2\, \frac{(m_s^+)^2}{M}\, N_+^\nu 
\Big)~.
\end{eqnarray}
Here we quote only the results for $\Phi_+$ components; the extension
to $\Phi_-$ follows upon some simple substitutions. The requirement of
exact SUSY at UV scale $M$ translates into the RGE boundary
conditions:
\( 
\widetilde{A}^\mu \Bigr|_M = \widetilde{B}^\mu \Bigr|_M = 0~.
\) 
In addition, the full set of equations include the RGE's for the soft breaking masses, 
\begin{eqnarray}
       \mu \frac{d m_s^2}
               {d\mu}            ~ =~ 
        \frac{\alpha}{4\pi}~ m_s^2~,
\end{eqnarray}
and RGE's for the dimension five SUSY LV operators
(\ref{RG_eqn_undiag}).   

Exact solutions of these RGE's are not
warranted for our purposes. Instead, we use the same approximation as
in (\ref{simplified}) together with  
$ \alpha/\pi \log (M/m_s) < 1$,
to obtain the solution 
\begin{equation}
\label{B_mu_coef}
\widetilde{B}^{\pm\nu} (m_s) ~=~ \frac{\alpha}{\pi}\, 
\log ( M/m_s )\,
\frac{(m_{s}^\pm)^2 }{M}\, 
\Big\{ 
\frac{3}
     {2} N^\nu(M) 
~-~ N_\pm^{\,\nu}(M)
\Big\}~,
\end{equation}
for $\widetilde{B}^\pm_\mu$ in the leading 
$\alpha\log$ approximation.

\subsection{Operators in the gauge sector. Chern-Simons term.}
\label{SB_gauge_sector}

The absence of optical activity effects caused by the CS term has been
checked over cosmological distances, providing 
a very sensitive probe of $k_\mu$ in (\ref{LVqed}) (see {\em e.g.}\ 
Ref.~\cite{CFJ} and references therein). The limit on $k_\mu$ is about
the present Hubble expansion rate, and is ten orders of magnitude
better than the level of sensitivity for the best terrestrial
experiments searching for LV parameters in (\ref{LVqed}). Not
surprisingly, the issue of CS term generated by radiative corrections
from other LV interactions has drawn a lot of interest
\cite{CG,Jackiw:1999yp,Chung:1998jv,Andrianov:2001zj,Perez-Victoria:2001ej},
exhibiting the whole range of answers for $k_\mu$ (including zero)
being induced by $b_\mu$. A no-go theorem by Coleman and Glashow
\cite{CG} indicates the absence of the radiatively generated CS
term. If suitably rephrased, it states that the CS term
cannot be induced to first order by gauge invariant LV interactions. 
In section \ref{noAnomaly} we have extended this theorem to the exact
SUSY LV interactions.

We would like to argue that below the soft SUSY breaking scale the CS term also 
cannot be generated. Indeed, the CS interaction can only be generated
by a fermion running in the loop, as a bosonic loop cannot produce 
$ \epsilon^{\mu\nu\rho\sigma} $ entering the expression for the CS term. 
However, the SUSY breaking terms (\ref{SB_vertex}) only 
provide masses for the bosonic components of chiral superfields
and thus only affect the scalar parts of the diagrams, which are 
incapable of inducing the CS interaction. (The possibility of a
soft gaugino mass is not relevant because diagrams that could induce
the CS term only include chiral matter fermions, not gauginos.) 
In particular, the no-go theorem by Coleman and Glashow \cite{CG} for
QED is re-obtained by sending the soft masses to infinity.

\begin{figure}
\begin{center}
\begin{tabular}{ccccc}
\includegraphics[width=2.7cm,height=2.7cm,keepaspectratio]
 {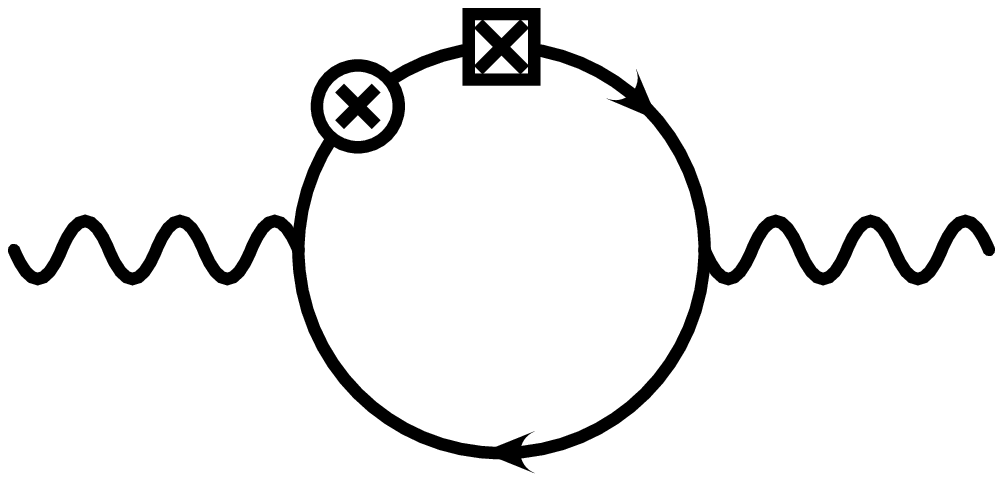} &
\includegraphics[width=2.7cm,height=2.7cm,keepaspectratio]
 {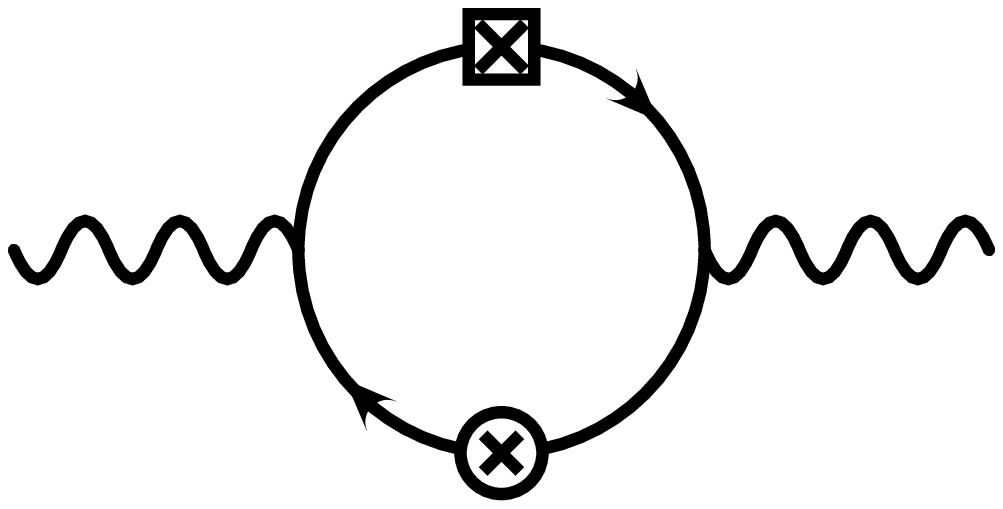} 
\includegraphics[width=2.7cm,height=2.7cm,keepaspectratio]
 {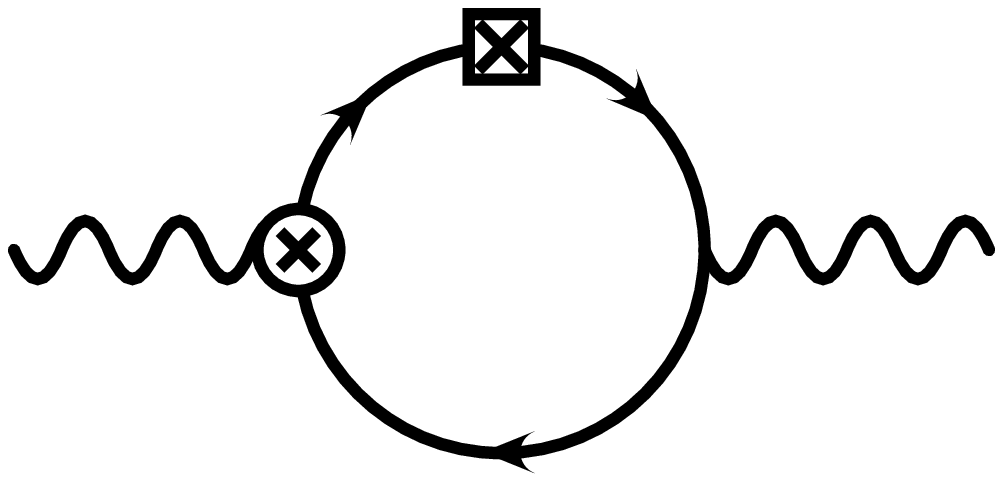} &
\includegraphics[width=2.7cm,height=2.7cm,keepaspectratio]
 {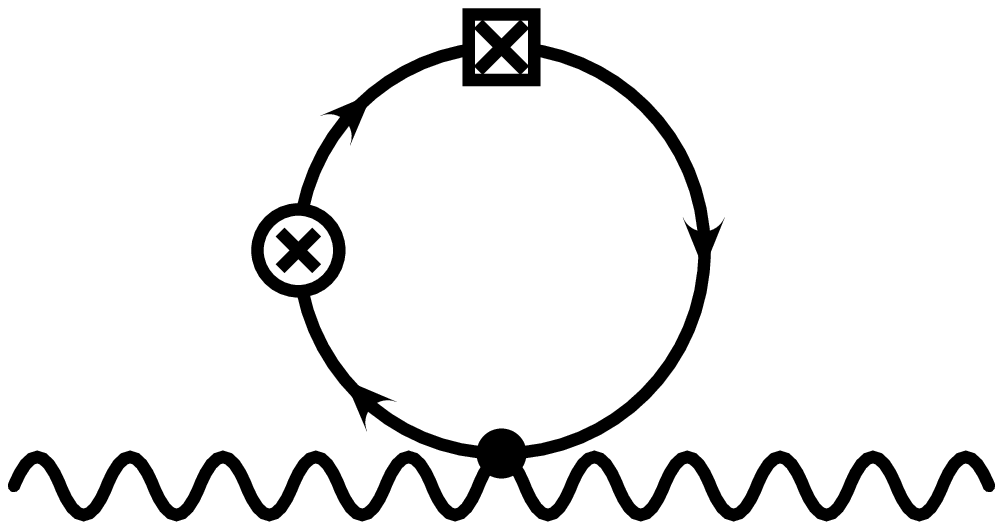} &
\includegraphics[width=2.7cm,height=2.7cm,keepaspectratio]
 {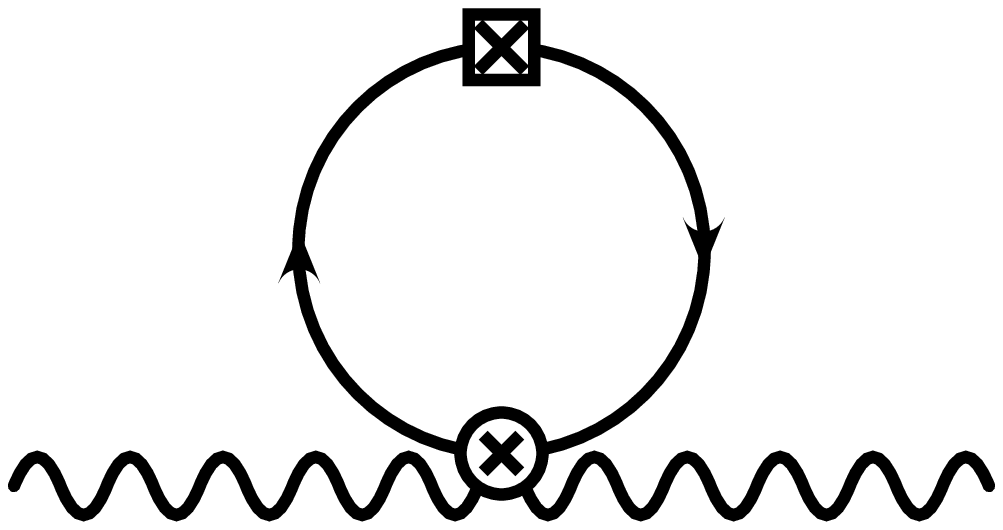} 
\end{tabular}
\end{center}
 \caption{\label{diag_SB_gauge}
Dimension three one-loop contributions arising from the
dimension five LV operators (\ref{LV_matter}) and 
soft supersymmetry breaking.
}
\end{figure}

We have confirmed this result by a direct calculation in the presence of
soft-breaking. The relevant diagrams, given in
Fig.~\ref{diag_SB_gauge}, are obtained by inserting the soft breaking
interaction (\ref{SB_vertex}) into the diagrams shown in
Fig.~\ref{diag_LV_gauge}. As for the direct confirmation of the
absence of the SUSY CS term in section \ref{noAnomaly}, instead of
calculating all possible terms, we have only concentrated on those
structures that can induce the CS term. Here, again, the vertex
cancellation property \eqref{ResummedPropVer} can be used quite
effectively to mutually cancel contributions of particular diagrams.
A straightforward calculation shows that all terms proportional to the
CS interaction indeed cancel.

Note that this statement is only valid for the pure CS term
$ \epsilon^{\mu\nu\rho\sigma}\, A_\nu \partial_\rho A_\sigma $, 
while there is no evidence against the other possible operator in the 
photon sector,  $ \lambda \sigma^\mu \overline{\lambda} $. 
However, the presence/absence of the latter term 
is obviously not very relevant for phenomenological applications.

\section{Phenomenology of LV SQED: LV observables and experimental limits}
\label{Phenomenology}
\subsection{Component Expressions for LV operators}

In order to derive phenomenological consequences of 
the LV operators, we need to obtain their component expressions.
First we consider the matter operators (\ref{LV_matter}).
The component form for the electron part is given by: 
\begin{gather}
\nonumber
  \mathcal{L}_{\mathrm{LV~dim~5}}^{\mathrm{matter\,(+)}} 
~ =~ \frac{N_+^\mu}{M} \Big[~
    i \bar{F}_+ \mathcal{D}_\mu F_+ ~+~
    i e \bar{z}_+ D \mathcal{D}_\mu z_+ ~-~
    i e \mathcal{D}_\mu(\bar{z}_+) D z_+ 
~+~ 
  \frac{1}{2}\bar{\psi}_+\mathcal{D}_{(\mu}\mathcal{D}_{\nu)}
               \bar{\sigma}^\nu \psi_+ 
\nonumber \\
  + ~ 
    i e \frac{\sqrt{2}}{2} \Big\{
               \overline{\psi}_+\bar\sigma_\mu\lambda F_+ 
       ~-~
               \overline{F}_+\overline{\lambda} \bar\sigma_\mu \psi_+
                         \Big\}  ~+~
    e^2 \bar{z}_+ \Big\{
               \lambda\sigma_\mu\bar{\lambda} 
       ~-~
               \overline{\lambda}\bar\sigma_\mu\lambda 
                       \Big\} z_+ 
~+~ 
    \frac{1}{2} e \overline{\psi}_+\bar\sigma_\mu D\psi_+
\nonumber \\
\nonumber
 -~ 
   \sqrt{2} e \Big\{ 
                     \mathcal{D}_\mu(\overline{\psi}_+)\overline{\lambda} z_+ 
     ~+~ 
                     \bar{z}_+ \lambda \mathcal{D}_\mu \psi_+ 
                     \Big\} 
    ~-~ 
    \frac{\sqrt{2}}{2} e \Big\{ 
                      \overline{\psi}_+\bar\sigma^\nu\sigma_\mu 
                     \bar{\lambda}\mathcal{D}_\nu z_+ +
                     \mathcal{D}_\nu(\bar{z}_+)\lambda\sigma_\mu
                     \bar{\sigma}^\nu \psi_+
                     \Big\}
   \\
  ~ -~
  \frac{1}{4} e \bar{\psi}_+\epsilon_\mu{}^{\nu\rho\sigma}
              F_{\rho\sigma} \bar{\sigma}_\nu \psi_+
   ~+~
  i \bar{z_+} \mathcal{D}^\nu \mathcal{D}_\mu \mathcal{D}_\nu z_+ 
   ~+~
   \frac{1}{2} i e \mathcal{D}_\nu (\bar{z}_+) \epsilon_\mu{}^{\nu\rho\sigma}
              F_{\rho\sigma} z_+ \, 
   \Big] ~,
\label{LV_electron_comp}
\end{gather}
This component representation  of dimension 5 LV operators allows 
for further
reduction of several terms in (\ref{LV_electron_comp}) on the
equations of motion. The result for this tedious but routine reduction
is given in Appendix~\ref{app_reduction}. To facilitate
phenomenological applications, we convert all Weyl spinors into
Dirac/Majorana four component spinors: 
\begin{equation}
\label{Dirac_spinors}
   \Psi ~=~
                 \begin{pmatrix}{c}
                    \psi_+ \\
                    \psi_-
                 \end{pmatrix}
          ~, \qquad  {\rm and} \qquad 
   \lambda ~=~ 
                 \begin{pmatrix}{c}
                    \lambda \\
                    \overline{\lambda}
                 \end{pmatrix}
~.
\end{equation}
Using notations (\ref{def_Nmu}) and (\ref{deltam}), we present the
dimension five LV operators (\ref{LV_matter}), containing electron and
photon fields, as:
\begin{equation}
   \mathcal{L}_{\rm LV}^{\rm matter} ~=~ 
        -
       \frac{N_A^\mu}
              {M} \, \frac{1}{2}e \,
       \overline{\Psi} \widetilde{F}_{\mu\nu}
                       \gamma_\nu \Psi 
     ~-~
        \frac{N_V^\mu}
              {M} \, \frac{1}{2}e \,
       \overline{\Psi} \widetilde{F}_{\mu\nu}
                       \gamma^\nu \gamma^5 \Psi 
~+~  \frac{N_V^\mu}{M} \,m_e^2\, \overline{\Psi} \gamma_\mu \Psi
     ~.
\label{resolved_LV_Dirac}
\end{equation}
Using the same notation, the dimension three operators 
(\ref{B_mu_coef}) can be rewritten as a vector and 
axial-vector operators:
%
\begin{eqnarray}
\nonumber
        \mathcal{L}_{\rm{LV\ dim\ 3}}^{\rm matter} &~=~ &
-\, \overline{\Psi} \gamma^\mu \Psi 
\Big\{\,
         m_s^2\, N_V^\mu 
~+~
 \frac{\Delta m^2}{2}\, N_A^\mu 
~-~
\frac{3}{2}\, \frac{\Delta m^2}{2}\, N^\mu
       \,
\Big\}\, \frac{\alpha\log (M/m_s) }{\pi M}
\\
\label{LV_induced_dim3}
&&
-\, \overline{\Psi} \gamma^\mu \gamma^5 \Psi 
\left\{\,
        m_s^2\, N_A^\mu 
~+~
\frac{\Delta m^2}{2}\, N_V^\mu 
~-~
\frac{3}{2}\, m_s^2\, N^\mu
       \,
\right\}\, \frac{\alpha\log (M/m_s)}{\pi M}~.
\end{eqnarray}
The next operator to consider in components is the photon operator
(\ref{LV_gauge}):
\begin{equation}
\label{LV_gauge_comp_again}
\mathcal{L}_{\mathrm{LV~dim~5}}^{\mathrm{gauge\ (K)}} 
 ~=~ \frac{N^\mu}{M}\,  
\Big( 
2\, \overline{\lambda}\,\gamma_\mu\, \Box\, 
   \lambda 
~+~
2\, \lambda\, \partial_\mu \slashed{\partial}\, 
   \overline{\lambda} 
~-~ 
2\, D\, \partial^\nu F_{\mu\nu}
~+~ 
\partial_\lambda F^{\lambda\nu}\, 
\widetilde{F}_{\nu\mu} 
\Big)~.
\end{equation}
These components are reducible on the equations of motion, 
and only the last term in (\ref{LV_gauge_comp_again})
leads to a contribution in the electron and photon sectors. 
By substituting $ \partial_\lambda F^{\lambda\mu} $ 
with the electromagnetic current 
$ J_{EM}^\mu=- e\, \overline{\Psi}\, \gamma^\mu \Psi $,
we get an interaction term
\begin{equation}
\label{LV_induced_by_gauge_K}
        \mathcal{L}_{\rm gauge\ (K)}^{\rm EOM} ~=~  
 e\, \overline{\Psi}\, N^\mu \gamma^\nu
\widetilde{F}_{\mu\nu}\, \Psi~,
\end{equation}
which  has the same form as the second term in (\ref{resolved_LV_Dirac}).
Notice, however, that this coincidence holds only within QED, 
as in the full MSSM $J_{EM}^\mu$ will also have other 
({\em i.e.}\ hadronic) contributions. Finally, the tensor operator
(\ref{LV_gauge_Tterm}) has the following component expression 
\begin{equation}
\mathcal{L}_{\mathrm{LV~dim~5}}^{\mathrm{gauge\ (T)}}  
      ~  =~ 
\frac{2T_{\mu\nu\rho}}{M}
\Big( 
F^{\nu\lambda}\partial^\mu F^\rho_{\phantom{\rho}\lambda}
\,-\, D\, \partial^\mu \widetilde{F}^{\nu\rho} 
\,-\,\overline{\lambda}\, \partial^\mu \partial_\sigma
\overline{\sigma}_\tau \lambda\; \epsilon^{\sigma\tau\nu\rho}
\Big)~.
\label{LV_gauge_Tterm_comp_again}
\end{equation}
It can be reduced on the equations of motion by using integration by
part and Jacobi identities.     
Applying the equations of motion to the pure
electromagnetic field strength term in 
(\ref{LV_gauge_Tterm_comp_again}), we obtain in the
electron-photon sector of QED
\begin{eqnarray}
        \mathcal{L}_{\rm gauge\ (T)}^{\rm EOM} ~=~   
\label{LV_induced_by_gauge_T}
        2\,e\,  T_{\mu\nu\rho} \, 
         \overline{\Psi} \gamma^\mu F^{\nu\rho} \Psi
        ~.
\end{eqnarray}
Confirming the general conclusion of \cite{GrootNibbelink:2004za}, we
observe that none of the LV operators give corrections to the EOM that
grow at high energies.

Now we gather all operators of phenomenological
interest of dimensions five and three, 
(\ref{LV_induced_dim3}), (\ref{resolved_LV_Dirac}),
(\ref{LV_induced_by_gauge_K}) and
(\ref{LV_induced_by_gauge_T}), in a single expression:
\begin{equation}
\label{L_eff}
 - \mathcal{L}_{\rm eff~LV}
         ~=~ 
\overline{\Psi}\, \gamma^\mu \Big( 
 a_\mu ~+~ b_\mu \gamma^5 
~+~ e\,c_\nu\,   \widetilde{F}^{\nu\mu} 
~+~  e\, d_\nu\,\widetilde{F}^{\nu\mu} \gamma^5
~+~e\, f_{\mu\rho\sigma}\,  F^{\rho\sigma} 
\Big) \Psi~, 
\end{equation} 
where we use the notations of \cite{Colladay:1998fq} for the
coefficients of dimension three operators. The Wilson coefficients in
(\ref{L_eff}) are expressed in terms of the original LV parameters,
electromagnetic coupling constant, and soft breaking masses: 
\begin{eqnarray}
\nonumber
        a^\mu & ~=~ &
         -\, \frac{1}{M}\, m_e^2 N_V^\mu 
        ~+~
        \frac{\alpha\, \log (M/m_s)}{\pi M}\, 
        \Big\{
         m_s^2\, N_V^\mu 
                ~+~
                 \frac{\Delta m^2}{2}\, 
                                             N_A^\mu
                ~-~
                \frac{3}{2}\, \frac{\Delta m^2}{2}\, 
                                               N^\mu  
        \Big\}~ ,
\\
\label{L_eff_coefs}
        b^\mu & ~=~ & 
        \frac{\alpha\,\log (M/m_s)}{\pi M} \, 
        \Big\{
                 m_s^2\, N_A^\mu
                ~+~
                 \frac{\Delta m^2}{2}\, 
                                             N_V^\mu
                ~-~
                \frac{3}{2}\, m_s^2\, N^\mu
        \Big\} ~,
\\
\nonumber
        c^\mu & ~=~ &
         \frac{1}{M}
        \Big\{ 
                \frac{1}{2}N_A^\mu
                ~-~
                N^\mu
        \Big\}~,\qquad 
        d^\mu ~=~
        \frac{1}{M}\, \frac{N_V^\mu}{2} 
~,\qquad 
        f^{\mu\nu\rho} ~ = ~
        \frac{2}{M} T^{\mu\,\nu\rho}
~.
\end{eqnarray}
The result is given in the leading $\alpha \log$ approximation, and with
all LV parameters normalized at the SUSY threshold
$m_s$. The operator proportional to $ a^\mu $ does not lead to 
any physical effects as it can be totally absorbed into
the kinetic term $ - i\, \overline{\Psi}\slashed{\partial}\Psi $
via a phase redefinition, $ \Psi(x) \to e^{i a^\mu x_\mu} \Psi(x) $.
The rest of the operators lead to observable LV signatures.

\subsection{Constraints on LV parameters of SQED}

Now, we are prepared to extract observational consequences of LV SQED,
and to impose constraints on the coefficients of the effective low-energy
Lagrangian (\ref{L_eff}). To do that, we derive the non-relativistic
effective Hamiltonian corresponding to (\ref{L_eff}) by splitting the
external backgrounds into spatial and  time-like components: 
\begin{eqnarray}
        \mathcal{H}_{\rm eff} 
        & ~=~ &
        \frac{\vec{p} \cdot \vec{a}}
                  {m}
        ~+~
        \vec{b} \cdot \vec{\sigma}
        ~+~
        \Big\{\, 
                \frac{e\, \vec{p}}
                    {m}
                \, ,\, 
                \big[ \vec{c} \times \vec{E} \big]
                ~-~
                c^0\, \vec{B} 
        \Big\}
        ~-~ 
        e \, d^0 \, \big( \vec{B} \cdot \vec{\sigma} \big)
\nonumber        \\
\label{H_eff}
        & &
        ~+~
        e\, \vec{d} \cdot
        \big[ \vec{E} \times \vec{\sigma} \big]
        ~+~
        \Big\{\, 
                \frac{e\, p^k}
                    {m}
                \,,\, 
                2\, f_{k0l}\, E^l 
                ~+~
                f_{klm} \,\epsilon_{lmn}\, 
                B^n
        \Big\}~. 
\end{eqnarray}
Here $ \vec{p} = - i \hbar \vec{\partial} $ is the momentum operator
and $\{.,.\}$ denotes the anti-commutator.

The tightest constraints come from the experiments searching for 
abnormal spin precession around external directions determined by the
LV vectors (\ref{L_eff_coefs}). These experiments limit LV parameters 
for electrons and nucleons. The relevant parameter we should compare 
our estimates with is the energy shift due to LV effects 
$\Delta \omega_{LV}$. In LV SQED, the effects are mediated by
dimension five operators, and therefore the strength of the
constraints on combinations of $M$ and LV backgrounds depends very
sensitively on the energy scale $\mu$ relating $\Delta \omega_{LV}$
and $M^{-1}$, $\Delta \omega_{LV} \sim \mu^2 M^{-1}$. Our analysis
shows that several possibilities for $\mu$ are possible: the soft
breaking scale, the hadronic scale ({\em i.e.}\ $\Lambda_{QCD}$), electron
mass, and finally, the energy scale given by an external magnetic or
electric field.

\subsubsection*{LV electron spin precession}

The soft  breaking scale enters in the LV parameter $b_\mu$, which is 
limited by torsion balance experiments searching for LV in the
electron sector \cite{Heckel:1999sy}. The sensitivity to the spatial
part of the axial-vector  
coupling $b^i$ achieved in this experiments is at the level better
than $|b^i| < 10^{-28}$ GeV. This condition imposes a stringent
constraint on the combination of the soft breaking masses, 
$M$, and LV parameters:
\begin{equation}
\frac{m_{s}^2}{(100~{\rm GeV})^2}~
\frac {10^{19}~{\rm GeV}}{M} ~ 
\left| N_A^i - \frac{3}{2} N^i + \delta_{s} N_V^i \right |
~<~10^{-12}~, 
\label{b_limit}
\end{equation}
where we normalize $M$ to the Planck scale. We have introduced a
dimensionless quantity  
$\delta_s = \Delta m^2/(2m_s^2)$ that parameterizes parity violation
in the soft breaking sector. The lightest values for $m_s$ not
excluded by direct collider searches are slightly above 
$100~{\rm GeV}$, and therefore $|N_A^i - \frac{3}{2} N^i |$ is limited
to be less than $10^{-12}$.

\subsubsection*{LV nuclear spin precession}

The next constraint uses the energy scale $\mu \sim \Lambda_{QCD}$ from
hadron physics. In order to obtain it, we have to go beyond pure QED and
include hadronic components in $J^\mu_{EM}$
(\ref{LV_induced_by_gauge_K}). Then, as discussed earlier in  
\cite{GrootNibbelink:2004za}, the LV SQED operator (\ref{LV_gauge})
gives rise to interaction of the spatial components of $N^\mu$,
an electric field, and the spatial component  of the hadronic current, 
${\cal L} = M^{-1}\epsilon_{ijk} N^i E^j J_{EM}^k$.
The average of such interaction inside the nucleus with spin $I$ leads to 
the effective Hamiltonian 
$\mathcal{H}_{\rm eff}= \kappa \vec I\cdot \vec N$.
The strength $\kappa$ of this interaction is given by a nuclear matrix
element, which can be estimated as the product of the typical value of 
the electric field inside a heavy nucleus times the characteristic  
nucleon momentum. Combined with a $10^{-32}$ GeV level of sensitivity
for  $\Delta \omega_{LV}$ in the most advanced experiments
\cite{clock1,clock2}, this results in a stringent bound on $|N^i|$:
\begin{equation}
\kappa ~\sim~ \frac{e\, E \, p_{\rm nucl}}{M\, m_p} 
~\sim~ \frac{Z^{1/3}\, {\rm fm}^{-3}\alpha}{M \,m_p} 
\qquad \Rightarrow \qquad
\frac {10^{19}~{\rm GeV}}{M} ~ |N^i| ~<~ 10^{-9}.
\label{N_limit_nucl}
\end{equation}
Here $m_p$ is the proton mass and $p_{\rm nucl}\sim {\rm fm}^{-1}$ is
the typical nucleon momentum. A more refined nuclear calculation can
be done for mercury and xenon nuclei used in \cite{clock1,clock2} if needed.

\subsubsection*{LV precession of the angular momentum of a paramagnetic atom}

If for some unexpected reasons the effective electron LV coupling
$b^\mu$ is close to  zero, the interaction term proportional to
$c^\mu$ in (\ref{H_eff}) would still induce a coupling of the electron
angular momentum $j$ inside a paramagnetic atom to the spatial
component of   
$N_A^\mu/2-N^\mu$ with $\mathcal{H}_{\rm eff}= \kappa j_i (N^i_A/2-N^i) $. 
In this case, the characteristic scale connecting 
$\Delta \omega_{LV}$ with LV parameters is the typical momentum 
of atomic electrons, $\mu \sim p_{\rm atomic} \sim \alpha m_e$. 
Apart from an overall coefficient, the atomic matrix element
 responsible for this interaction has the same strength as the usual spin-orbit 
 interaction, resulting in the estimate of $\kappa$
 \begin{equation}
 \kappa ~\sim~ Z^2\alpha^2 \, \frac{\alpha^2 \, m_e^2}{M}
\qquad \Rightarrow \qquad
\frac {10^{19}~{\rm GeV}}{M} ~ |N^i-N_A^i/2|~<~10^{-2}.
 \label{N_limit_atom}
 \end{equation}

\subsubsection*{CPT-odd anomalous magnetic moment of electrons and positrons}

The limits explored so far do not use the fact that LV operators of
dimension five break CPT, whereas the experiments \cite{clock1,clock2} are
done with normal matter. Some other experiments explicitly compare
properties  of electrons and positrons, and can therefore be used 
to constrain LV CPT-odd operators. For example, a $ d^0 $-proportional
term in (\ref{H_eff}) induces an interaction between the electron spin
with a magnetic field, and thus contributes to the anomalous magnetic
moment of electrons and positrons. The different $g$-factors for
electrons and positrons are limited at $10^{-12}$ level: 
$|g_{e}-g_{\bar e}|< 8\times 10^{-12}$ \cite{Mittleman:1999it}.  

The interaction Hamiltonian for electrons and positrons, corrected by the 
CPT-odd $d^0$-proportional interaction, takes the form: 
\begin{equation}
 \mathcal{H}_{\rm eff}^{e} ~=~ 
-\, e\, d^0\, \frac{\vec{B}\cdot\vec{S}}{S}
~-~ 
|\mu|\, \frac{\vec{B}\cdot\vec{S}}{S}~, 
\qquad 
 \mathcal{H}_{\rm eff}^{\bar{e}} ~=~  
-\, e\, d^0\, \frac{\vec{B}\cdot\vec{S}}{S}
~+~ 
|\mu|\, \frac{\vec{B}\cdot\vec{S}}{S}~.
\end{equation}
This gives a bound on the time-like component of $N_V^\mu$: 
\begin{equation}
 \frac{m_e}{M}\,  |N_V^0| ~<~ 2\times 10^{-12} 
\qquad \Rightarrow \qquad 
\frac {10^{19}~{\rm GeV}}{M} ~ |N_V^0| ~<~ 10^{10} 
~.
\end{equation}
Obviously, this limit is inferior to those derived from searches of
the breakdown of rotational invariance \cite{clock1,clock2}.

It is interesting to note that the CPT-violating correction to
the magnetic moments of electrons and positrons arises in LV SQED 
even when SUSY is unbroken. At first sight this seems to 
be at odds with the Ferrara-Remiddi theorem which
forbids emergence of the anomalous magnetic moment of the
electron in the exact SUSY limit \cite{Ferrara:1974wb}: 
the anomalous magnetic moment of the electron, 
$e(4m)^{-1} \overline{\Psi} \sigma_{\mu\nu} \Psi \; F^{\mu\nu}$,
should appear as the highest component of some superfield, but
no such supermultiplet exists \cite{Ferrara:1974wb}. However, one of
the assumptions leading to this result is Lorentz invariance, therefore 
when SQED is extended by LV operators, the anomalous magnetic 
moment, 
\( 
eM^{-1}N_-^\mu \, \overline{\Psi} \widetilde{F}_{\mu\nu}
\gamma^\nu \gamma^5 \Psi~, 
\) 
does arise as the highest component of a superfield  operator, namely 
(\ref{LV_matter}).

\subsubsection*{Consequences for some dimension five operators in LV
SQED}

The two most stringent limits, (\ref{b_limit}) and (\ref{N_limit_nucl}),
are sensitive to different linear combinations of $N^i$ and $N_A^i$ vectors,
thus imposing similar strength constraints on $N^i$ and $N_A^i$
separately. In order to impose a constraint on $N_V^i$, one has to  
make further assumptions about $\delta_s$. In the full MSSM scenario
(as opposed to its SQED subset), parity is broken above the weak
scale. 
Hence,  a $\delta_s $ at a percent level or larger
arises form radiative corrections even if the boundary conditions at
$M$ are parity conserving, {\em i.e.}\ $m_s^+(M) = m_s^-(M)$. 
This provides a sensitivity to $N_V^i$ at the level of $10^{-10}$. 
The time-like components of vectors $N_\pm^\mu$ and $N^\mu $ are 
also constrained:  
the motion of the earth and the solar system introduces a dependence
of the laboratory frame  on the velocity relative to the fixed vector
backgrounds. Therefore, a non-zero $N^0$ would "mix" with 
$N^i$ at $O(v/c) \sim 10^{-3}-10^{-4}$ level. As a result, 
$10^{-6}-10^{-8}$ level constraints can be 
imposed on $N_V^0$, $N_A^0$ and $N^0$.

LV induced by $ T^{\mu\,\nu\rho} $ (\ref{LV_gauge_Tterm})
is also subject to experimental constraints. For example, 
a three dimensional vector $ f^k=\epsilon_{ijk} T^{i\,0j}$, 
obtained from the tensor $T^{\mu\,\nu\rho} $, leads essentially to the
same effects as  vector $c^k$, and is therefore subject to bounds
analogous to (\ref{N_limit_nucl}) and (\ref{N_limit_atom}). Other
components of $ T^{\mu\,\nu\rho} $ can be limited using   
their contributions to $f^k$ caused by earth motion effects.

\subsubsection*{Absence of Planck-scale bounds on dimension six LV operators}

Finally, we would like to assert that limits on dimension six operators 
are {\em not} able to rule out LV modifications at $M^{-2}_{\rm Pl}$ level. 
Many of the operators listed in (\ref{LV_dim6_Fterm}) and
(\ref{LV_dim6_Dterm}) contain antisymmetric tensors.  After the
inclusion of SUSY breaking, such terms can mix with the
$m_e\, \bar \Psi \sigma_{\mu\nu} \Psi$ operator, leading to 
LV spin precession of the electron. Assuming that the sizes of the
dimensionless tensors in (\ref{LV_dim6_Fterm}) and
(\ref{LV_dim6_Dterm}) are $O(1)$, one can estimate the sensitivity to $M$
via the dimension six LV operators:  
$M^2 \sim m_em_s^2/(10^{-28}~{\rm GeV}). $
This translates into a bound of $M \sim 10^{14}$ GeV for $m_s \sim
100$ GeV, which is lower than the Planck scale. On the other hand, we
notice that in the framework of the LV MSSM the sensitivity to $M$ via
dimension six LV operators will be higher, when observables in the
quark sector are employed. Indeed, we expect $m_e$ to be replaced 
by $m_q$, and $10^{-28}~{\rm GeV}$ by $10^{-32}~{\rm GeV}$, as 
experiments searching for anomalous spin precession of nucleons are
intrinsically more precise.  In this case the sensitivity to $M$
would get close to the Planck scale, and future increase of the
experimental sensitivity  may probe this type of models. Although
undoubtedly very interesting, more detailed study of the observational
consequences of  dimension six LV operators goes beyond the scope of
the present paper.

\section{Discussion and Conclusions}
\label{conclusion}

We have constructed a dimension five LV extension of SQED, 
as a subset of the full LV MSSM. The LV modifications are 
power-suppressed by the UV scale $M$ and decouple in the limit of
$M\to\infty$.  In the leading order in the inverse UV scale, $O(M^{-1})$, 
dimension five LV operators can be coupled to two types of LV
backgrounds. There are three background vectors  
$N^\mu$, $N^\mu_+$ and $N^{\mu}_-$, as well as an irreducible rank
three tensor $T^{\mu\,\nu\lambda}$ (antisymmetric in $\nu\lambda$). The
corresponding LV operators are all CPT-odd. At the dimension six level
LV operators are CPT-even; their classification has been given in 
this paper.

We have explored quantum effects in the presence of the 
LV terms. We have shown that no $D$-term is induced and 
the anomaly cancellation condition is not altered by the presence of
LV in the limit of exact SUSY. The RGE's for LV operators of dimension
five were derived in the limits of exact and softly-broken SUSY. Once
SUSY is broken, dimension three operators can be generated.  
The transmutation of dimension five LV operators into dimension three
is controlled by the scale of soft SUSY breaking. This alleviates the 
LV naturalness problem, because the potentially problematic quadratic
divergences are stabilized at the SUSY breakdown scale.
In order to obtain phenomenologically applicable formulas, we broke
SUSY by introducing scalar electron masses, and calculated the resulting effective
LV Lagrangian for electrons. A dimension three
operator for photons, the CS term, is not generated at the
loop level. It is remarkable that none of the LV operators, considered
in this paper, lead to high-energy modifications of dispersion relations.  
Therefore, none of the stringent astrophysics-derived limits on LV
parameters  \cite{Ted1,GK} apply to LV SQED.

We have obtained explicit component expressions for LV
interactions generated by vector and tensor backgrounds, which allowed
us to derive observational consequences of LV in SQED. Using the results
of high-precision searches for LV spin interactions, we derived  
stringent limits on some linear combinations of LV parameters. The most
stringent results resulted from a one-loop induced coupling between the
electron axial vector current and some combination of the background
vectors $N^\mu$, $N^\mu_+$ and $N^{\mu}_-$. The strongest bound was
obtained from the absence of anomalous spin precession for electrons,
which is checked at a level better than $10^{-28}$ GeV by
torsion balance experiments \cite{Heckel:1999sy}. Assuming that the UV
scale is of the same order of magnitude as the Planck scale, 
we were able to constrain one linear combination of $N^i$, $N^i_+$ and
$N^i_-$ at the level better than $O(10^{-12})$. Conversely, if we
insist that $N^i \sim O(1)$,  such experiments provide a sensitivity to
the LV ultraviolet scale which is more than ten orders of magnitude
{\em larger} than the Planck scale. Other precision experiments
\cite{clock1,clock2} provide stringent constraints on different
linear combinations of the LV vector and tensor backgrounds.

The existence of strong constraints on LV at dimension five level
(with or without SUSY), poses a serious challenge for theories that
predict LV at $1/M_{\rm Pl}$ level. Therefore, either such theories are
ruled out, or they require abandoning an effective field theory
description of LV. (The latter does not seem a reasonable alternative
to us.) However, it might be that dimension five operators are
forbidden by some additional symmetry reasons, such as {\em e.g.}\
CPT. At the next order, $O(M^{-2})$, Planck suppressed LV effects 
are not excluded. (The best constraints may be better than the Planck
scale \cite{Gagnon:2004xh}, but are applicable only to operators that 
modify high-energy dispersion relations.) The classification of
dimension six LV operators in SQED has shown that they  couple to
symmetric or antisymmetric two-index tensor backgrounds. 
Non of these operators lead to modifications of the dispersion
relations hence the bounds of \cite{Gagnon:2004xh} do not apply. 
As we discussed at length for dimension five LV operators, similarly,  some
of dimension six operators will transmute into dimension four operators
due to quantum effects in the presence of soft-breaking
terms. The scale of the transmutation is controlled by the
SUSY breakdown scale, which gives an estimate for the size of  
LV backgrounds at dimension four as $m_s^2/M^2 \sim 10^{-32}$
for $m_s\sim1$ TeV and $M\sim 10^{19}$ GeV. This prediction 
comes close to the experimental sensitivity to such operators,
and therefore deserves further study in the framework of LV MSSM.

\section*{Acknowledgments}

SGN is indebted to M.\ Voloshin and M.\ Shifman for useful discussions
about related field theoretical issues. 
SGN would like to thank the Perimeter Institute and the University of
Guelph for their kind hospitality at various stages of this project. 
The work of P.B. and M.P. is supported in part by the N.S.E.R.C. of Canada. SGN has been supported in part by the Department of Energy 
under contract DE--FG02--94ER40823 at the University of Minnesota.

\pagebreak
\appendix

\section{Reduction of chiral LV operators on equations of motion}
\label{app_reduction}

The component expressions for LV terms in the chiral sector are given 
in (\ref{LV_electron_comp}). They can be transformed further 
by eliminating the auxiliary fields and higher derivatives 
via the equations of motion. Writing the result in terms of Dirac
four-spinors, we get the following rather lengthy expression:
\begin{eqnarray}
\nonumber
\lefteqn{
     \mathcal{L}_{\mathrm{LV}}^{\mathrm{matter}}  ~= 
~      
-~
\frac{N_A^\mu}{M}\,
\frac{1}{4} \,e\,
\overline{\Psi} \epsilon_{\mu\nu\rho\sigma}
F^{\rho\sigma} \gamma^\nu \Psi      
~-~
\frac{N_V^\mu}{M}\,
\frac{1}{4} \,e\,
\overline{\Psi} \epsilon_{\mu\nu\rho\sigma}
F^{\rho\sigma} \gamma^\nu \gamma^5 \Psi 
~+~ 
}\\
\nonumber
&&
~+~
\frac{N_+^\mu}{M}
\Big[\,
\frac{1}{2}i \,e \, 
\mathcal{D}^\nu \overline{z}_+ \,
\epsilon_{\mu\nu\rho\sigma}F^{\rho\sigma} z_+ 
~+~
\frac{1}{2}\, e\,
\Big(
  \overline{z}_+ F_{\mu\nu}
  \mathcal{D}^\nu z_+ 
  ~+~
  \mathcal{D}^\nu \overline{z}_+ \,
  F_{\mu\nu} z_+
\Big) 
\\
\nonumber
&&
               \qquad
~-~
\frac{i}{2}\, e^2\,
\Big(
  \mathcal{D}_\mu \overline{z}_+ \,
   z_+ 
  ~-~
  \overline{z}_+ \mathcal{D}_\mu z_+
\Big)
\Big\{
  z_-  \overline{z}_- 
  ~-~
  \overline{z}_+  z_+
\Big\}
\,\Big] ~+~ \\
\nonumber
&&
~+~
\frac{N_-^\mu}{M}
\Big[\,
- \frac{1}{2} i\, e\,
z_- \epsilon_{\mu\nu\rho\sigma}F^{\rho\sigma}
\mathcal{D}_\nu \overline{z}_- 
~-~
\frac{1}{2}\, e\,
\Big(
  z_- F_{\mu\nu}\mathcal{D}^\nu \overline{z}_- 
  ~+~
  \mathcal{D}^\nu z_- \,
  F_{\mu\nu} \overline{z}_- 
\Big)
\\
\nonumber
&&
               \qquad
~-~ 
\frac{i}{2}\, e^2\,
\Big(
  \mathcal{D}_\mu z_- \overline{z}_-
  ~-~
  z_- \mathcal{D}_\mu \overline{z}_-
\Big)
\Big\{
  z_- \overline{z}_- 
  ~-~
  \overline{z}_+ z_+
\Big\}
\,\Big]
~-~ \\
\nonumber
&&
~-~
\frac{N_+^\mu}{M}\, e^2\,
\overline{z}_+ \, \overline{\lambda} \gamma^\mu \gamma^5 
\lambda\, z_+ 
~-~
\frac{N_-^\mu}{M}\, e^2\,
z_-\, \overline{\lambda}\gamma^\mu\gamma^5
\lambda\, \overline{z}_-
~-~ \\
\nonumber
&&
~-~
\frac{N_{+\mu}}{M}\,
\frac{\sqrt{2}}{2}\, e\,
\Big(\,
\overline{\Psi} \gamma^\nu\gamma^\mu P_R
\lambda \, \mathcal{D}_\nu z_+ 
~+~
\mathcal{D}_\nu \overline{z}_+ \,
\overline{\lambda} \gamma^\mu \gamma^\nu
P_L \Psi
\,
\Big)
~+~ \\
\nonumber
&&
~+~
\frac{N_{-\mu}}{M}\,
\frac{\sqrt{2}}{2}\,e\,
\Big(\,
\mathcal{D}_\nu z_- \,
\overline{\lambda}\gamma^\mu\gamma^\nu P_R \Psi
~+~
\overline{\Psi}\gamma^\nu\gamma^\mu P_L \lambda
\, \mathcal{D}_\nu \overline{z}_-
\,
\Big)
~+~ \\
\label{LV_matter_component}
&&
~+~
\frac{N_+^\mu}{M}\,
\frac{\sqrt{2}}{2}\, e\,
\Big(\,
\overline{\Psi}P_R\, \mathcal{D}_\mu \lambda
\, z_+ 
~+~
\overline{z}_+ \,
\mathcal{D}_\mu 
\overline{\lambda}\; P_L \Psi
\,\Big)
~-~ \\
\nonumber
&&
~-~
\frac{N_-^\mu}{M}\,
\frac{\sqrt{2}}{2}\, e\,
\Big(\,
z_-\; \mathcal{D}_\mu \overline{\lambda} ~
P_R \Psi 
~+~
\overline{\Psi} P_L \, \mathcal{D}_\mu \lambda ~
\overline{z}_-
\,\Big) 
~+~ \\
\nonumber
&&
~+~ 
\frac{N_{A\mu}}{M}\,
\frac{1}{2}\, e^2\,
\overline{\Psi}\gamma^\mu  \Psi \,
\Big\{
  z_-  \overline{z}_- 
  ~-~
  \overline{z}_+  z_+
\Big\}
~+~ \\
\nonumber
&&
~+~
\frac{N_{V\mu}}{M}\,
\frac{1}{2}\, e^2\,
\overline{\Psi}\gamma^\mu \gamma^5 \Psi \,
\Big\{
  z_-  \overline{z}_- 
  ~-~
  \overline{z}_+  z_+
\Big\}
~-~ \\
\nonumber
&&
~-~
\frac{N_{A\mu}}{M}\,
\frac{\sqrt{2}}{2}\, i\, e\,
\Big(\,
\overline{m}_e\, \overline{\Psi} \gamma^\mu P_L
\lambda \overline{z}_- 
~-~
m_e\, z_- \overline{\lambda}
\gamma^\mu P_L \Psi
\,\Big)
~+~ \\
\nonumber
&&
~+~
\frac{N_{A\mu}}{M}\,
\frac{\sqrt{2}}{2}\, i\, e\,
\Big(\,
m_e\, \overline{\Psi}\gamma^\mu P_R \lambda\, z_+ 
~-~
\overline{m}_e\, \overline{z}_+\; \overline{\lambda}
\gamma^\mu P_R \Psi
\,\Big)
~+~ \\
\nonumber
&&
~+~ 
\frac{N_{V\mu}}{M}\, 2 i\, m_e \overline{m}_e \,
\big( 
\overline{z}_+ \mathcal{D}_\mu z_+ 
~+~
z_- \mathcal{D}_\mu \overline{z}_-
\big)~+~
\frac{N_{V\mu}}{M}\,
m_e \overline{m}_e \,
\overline{\Psi} \gamma^\mu \Psi
~.
\end{eqnarray}
The first, second, and last terms in \eqref{LV_matter_component}  enter 
the reduced Lagrangian (\ref{resolved_LV_Dirac}) that involves only
electrons, positrons and photons. The rather lengthy form of
\eqref{LV_matter_component} and the large number of diagrams that 
these interactions can create, underline the superiority of the 
superfield method over the component calculations for 
all processes with momenta larger than $m_s$.

\section{Conventions and notations}
\label{app_conventions}

Our notations for the superfield formalism are based on 
Wess \& Bagger \cite{Wess:1992cp}.
Covariant derivatives and hermitean conjugation are taken from
\cite{Gates:1983nr} with a proper adaptation. 
We use the  $ (-+++) $ metric signature.
All spinor algebra definitions can be found in 
\cite{Wess:1992cp},
and we list here only some minor conventional departures.
Unlike in \cite{Wess:1992cp}, we denote the space-time Lorentz
indices by letters from the middle of the \emph{Greek}
alphabet:
$ v_\mu $, $ \sigma_\nu $, $ N^\rho $, etc,
as one is normally accustomed to in QFT.
Spinor indices are taken, also as commonly accepted, from the
beginning of the Greek alphabet:
$ \theta^\alpha $, $ \epsilon_{\beta\gamma} $, 
$ \overline{\psi}_{\dot\delta}$.
Spinor derivatives are designated as
\[ 
\partial_\alpha ~=~ \frac{\partial}{\partial\theta^\alpha}~,
\qquad
\partial^\alpha ~=~ \epsilon^{\alpha\beta}\partial_\beta~.
\]

We use a notation with a slash in the case where a Lorentz
vector is contracted with a $ \sigma $-matrix, or a $ \gamma $-matrix:
\begin{equation}
\nonumber 
\slashed{v} ~=~ v^\mu\, \sigma_\mu~, 
\qquad
\overline{\slashed{A}} ~=~ A^\mu\, \overline{\sigma}_\mu~, 
\qquad
\slashed{n} ~=~ n^\mu \gamma_\mu~.
\end{equation}

For switching from Weyl to Dirac spinors we followed the notations of 
\cite{Martin:1997ns}.
Weyl representation for Dirac spinors is the most appropriate in this case,
where two Weyl spinors combine into one Dirac spinor:
\[
\Psi ~=~  
\left (
\begin{array}{c}
  \xi_\alpha \\
\overline{\chi}^{\dot\alpha}
\end{array}
\right )~,
\qquad
\overline{\Psi} ~=~  
\left (
\begin{array}{c}
  \chi^\alpha \\
\overline{\xi}_{\dot\alpha}
\end{array}
\right )~,
\]
and the $ \gamma $-matrices take the form
\begin{eqnarray*}
\gamma^\mu ~=~ 
\left ( 
\begin{array}{cc}
0                    &    \sigma^\mu \\
                     \overline{\sigma}^\mu   &         0    
\end{array}
\right )~,
\qquad
\gamma^5 ~=~ 
\left ( 
\begin{array}{cc}
1      &         0  \\
                        0      &        -1    
\end{array}
\right )~.
\end{eqnarray*}

For complex conjugation, we use the notion of 
\emph{hermitean conjugation} defined
in 
\cite{Gates:1983nr}.
When translated into the Wess \& Bagger notations, it implies
\renewcommand{\arraystretch}{1.3}
\begin{eqnarray*}
\begin{array}{ll}
( \psi_\alpha )^\dagger ~=~ \overline{\psi}_{\dot\alpha}~,
&
\qquad
( \psi^\alpha )^\dagger ~=~ \overline{\psi}^{\dot\alpha}~, 
\\
\partial_\alpha^\dagger ~=~ \overline{\partial}_{\dot\alpha}~,
&
\qquad
\partial_\mu^\dagger ~=~ -\, \partial_\mu~, 
\\
D_\alpha^\dagger ~=~ -\, \overline{D}_{\dot\alpha}~,
&
\qquad
( \nabla_\alpha )^\dagger ~=~ -\, 
\overline{\nabla}_{\dot\alpha}~, 
\\
W_\alpha^\dagger ~=~ \overline{W}_{\dot\alpha}~. 
\end{array}
\renewcommand{\arraystretch}{1.0}
\end{eqnarray*}

Finally, the expansion of the chiral superfields of SQED in components is
defined as
\[
        \Phi_\pm ~=~ z_\pm ~+~ \sqrt{2}\, \theta\psi_\pm ~+~ \theta^2 F_\pm~, 
\]
while the vector superfield in the Wess-Zumino gauge is given by
\[
        V ~=~  -~ \theta\,\sigma^\mu\, \overline{\theta}\, A_\mu ~+~
                i \theta^2\, \overline{\theta}\, \overline{\lambda} 
                ~-~
                i \overline{\theta}{}^2\, \theta\lambda
                ~+~
                \frac{1}{2}
                \theta^2\overline{\theta}{}^2\, D~.
\]

\bibliographystyle{apsrev}
\bibliography{lorentz}

\end{document}